\documentclass[bibyear]{aa}  
\usepackage{graphicx}
\usepackage[flushleft]{threeparttable}
\usepackage{txfonts}
\usepackage{diagbox}
\makeatletter
\renewcommand*\aa@pageof{, page \thepage{} of \pageref*{LastPage}}
\makeatother
\usepackage{natbib,twoopt}
\usepackage[breaklinks=true,linkcolor=blue,allcolors=blue,colorlinks=true]{hyperref} 
\bibpunct{(}{)}{;}{a}{}{,}             
\makeatletter
  \newcommandtwoopt{\citeads}[3][][]{\href{http://adsabs.harvard.edu/abs/#3}%
    {\def\hyper@linkstart##1##2{}%
     \let\hyper@linkend\@empty\citealp[#1][#2]{#3}}}
  \newcommandtwoopt{\citepads}[3][][]{\href{http://adsabs.harvard.edu/abs/#3}%
    {\def\hyper@linkstart##1##2{}%
     \let\hyper@linkend\@empty\citep[#1][#2]{#3}}}
  \newcommandtwoopt{\citetads}[3][][]{\href{http://adsabs.harvard.edu/abs/#3}%
    {\def\hyper@linkstart##1##2{}%
     \let\hyper@linkend\@empty\citet[#1][#2]{#3}}}
  \newcommandtwoopt{\citeyearads}[3][][]%
    {\href{http://adsabs.harvard.edu/abs/#3}
    {\def\hyper@linkstart##1##2{}%
     \let\hyper@linkend\@empty\citeyear[#1][#2]{#3}}}
\makeatother
\usepackage{orcidlink}

\usepackage{booktabs}
\usepackage{colortbl}

\begin{document} 

  \title{Star-formation rate and stellar mass calibrations based on infrared photometry and their dependence on stellar population age and extinction}
   \titlerunning{IR SFR and $M_\star$ calibrations, and their dependence on SP age and extinction}

   \author{K. Kouroumpatzakis
          \inst{1,}\inst{2,}\inst{3}\,\orcidlink{0000-0002-1444-2016}\fnmsep\thanks{email:konstantinos.kouroumpatzakis@asu.cas.cz}
          \and
          A. Zezas\inst{3,}\inst{4,}\inst{2}\,\orcidlink{0000-0001-8952-676X}
          \and
          E. Kyritsis\inst{3,}\inst{2}\,\orcidlink{0000-0003-1497-1134}
          \and 
          S. Salim\inst{5}
          \and 
          J. Svoboda\inst{1}\,\orcidlink{0000-0003-2931-0742}
          }

   \institute{
             Astronomical Institute, Academy of Sciences, Boční II 1401, CZ-14131 Prague, Czech Republic\\
              \email{konstantinos.kouroumpatzakis@asu.cas.cz}
              \and
              Institute of Astrophysics, FORTH, GR-71110 Heraklion, Greece
             \and
             Department of Physics, Univ. of Crete, GR-70013 Heraklion, Greece
             \and
             Center for Astrophysics \textbar\ Harvard \& Smithsonian, 60 Garden St., Cambridge, MA 02138, USA
             \and
             Department of Astronomy, Indiana University, Bloomington, IN 47404, USA\\
             }

   \date{Received February 01, 2023; accepted March 16, 2023}

 
  \abstract
   {The stellar mass ($M_\star$) and the star-formation rate (SFR) are among the most important features that characterize galaxies.
   Measuring these fundamental properties accurately is critical for understanding the present state of galaxies, their history, and future evolution.
   Infrared (IR) photometry is widely used to measure the $M_\star$ and SFR of galaxies because the near-IR traces the continuum emission of the bulk of their stellar populations (SPs), and the mid/far-IR traces the dust emission powered by star-forming activity.}
   {This work explores the dependence of the IR emission of galaxies on their extinction, and the age of their SPs.
   It aims at providing accurate and precise IR-photometry SFR and $M_\star$ calibrations that account for SP age and extinction while providing quantification of their scatter.}
   {We use the \texttt{CIGALE} spectral energy distribution (SED) fitting code to create model SEDs of galaxies with a wide range of star-formation histories, dust content, and interstellar medium properties.
   We fit the relations between $M_\star$ and SFR with IR and optical photometry of the model-galaxy SEDs with the Markov-chain Monte-Carlo (MCMC) method. 
   As an independent confirmation of the MCMC fitting method, we perform a machine-learning random forest (RF) analysis on the same data set.
   The RF model yields similar results to the MCMC fits validating the latter.}
   {This work provides calibrations for the SFR using a combination of the WISE bands 1 and 3, or the JWST NIR-F200W and MIRI-F2100W.
   It also provides mass-to-light ratio calibrations based on the WISE band-1, the JWST NIR-F200W, and the optical $u{-}r$ or $g{-}r$ colors.
   These calibrations account for the biases attributed to the SP age, while they are given in the form of extinction-dependent and extinction-independent relations.
   }
   {The proposed calibrations show robust estimations while minimizing the scatter and biases throughout a wide range of SFRs and stellar masses.
   The SFR calibration offers better results, especially in dust-free or passive galaxies where the contributions of old SPs or biases from the lack of dust are significant.
   Similarly, the $M_\star$ calibration yields significantly better results for dusty/high-SFR galaxies where dust emission can otherwise bias the estimations.}


   \maketitle
%
\section{Introduction}
\label{sec:intro}

Galaxies are among the most important structures of the Universe hosting the bulk of its cold baryonic mass.
The fundamental process of star formation (SF) which takes place within galaxies, transforms the gas into stars and therefore, plays a significant role in all aspects of galaxy evolution, and their current state.
For example, the current star-formation rate (SFR) of galaxies is strongly correlated with their stellar mass \citep[$M_\star$; e.g.][a.k.a. \textit{main sequence of galaxies}]{2007A&A...468...33E}, with their galactic winds \citep[e.g.][]{1990ApJS...74..833H,1996ApJ...462..651L}, the number of supernovae \citep[e.g.][]{2005ApJ...626..864M}, and their X-ray luminosity originating from high-mass X-ray binaries \citep[e.g.][]{2012MNRAS.419.2095M,2020MNRAS.494.5967K}.

Similarly, the $M_\star$ of galaxies is strongly correlated with their kinematic and morphological parameters \citep[e.g.][]{2005ApJ...625..621B,2006Natur.442..786G} while it is also related to the nature of their stellar content and the properties of their interstellar medium (ISM), for example through the mass--metallicity relation \citep[e.g.][]{2004ApJ...613..898T}.
Moreover, there is a strong correlation between the $M_\star$ of the bulges of galaxies with the mass of their central supermassive black hole \citep[e.g.][]{2013ARA&A..51..511K}.
Therefore, being able to accurately estimate the SFR and $M_\star$ is crucial for most studies involving galaxies, and studies of the evolution of the Universe.

Galaxies come in many forms and shapes, and in different states with respect to their current star-forming activity or ISM properties.
This is the main reason for discrepancies in measuring the SFR because the various SFR tracers are based on different emission mechanisms related to SF. 
These tracers include: the ultraviolet (UV) light coming directly from the photospheres of massive stars, emission lines from nebulae that have been ionized by star-forming activity (e.g. H$\alpha$, [\ion{O}{III}]), infrared (IR) emission from dust that was heated by the young stars' UV and optical emission, radio emission from relativistic electrons related to supernovae activity, and several others
\citep[for a review see][]{2012ARA&A..50..531K}.

Depending on the availability of observations different methods to estimate the SFR are used.
However, it has been shown that there can be significant discrepancies between these methods, sometimes up to an order of magnitude \citep[e.g.][]{2012ARA&A..50..531K}.
In particular, although monochromatic IR-based SFR indicators are good for dusty/high-SFR galaxies, they can lead to significant underestimation of the true SFR in low-mass or low-metallicity dust-deficient galaxies \citep[e.g.][]{2007ApJ...666..870C,2021MNRAS.506.3079K}.
Moreover, IR SFR tracers can be also biased by the age of the stellar population \citep[SP; e.g.][]{2008MNRAS.386.1157C,2012AJ....144....3L,2014A&A...571A..72B,2016A&A...589A.108C,2019A&A...624A..80N,2020MNRAS.494.5967K} or stochastic heating of the dust
\citep[e.g.][]{2015A&A...580A..87C,2019A&A...631A..38L}.
On the other hand, the use of UV or H$\alpha$ emission as SFR tracers require precise measurements of extinction in order to provide reliable estimations.

A possible solution is to use one of the \textit{hybrid} SFR indicators \citep[e.g. 24$\mu$m~+~H$\alpha$, FIR~+~UV; e.g.][]{2009ApJ...703.1672K, 2011ApJ...741..124H}.
They combine emission that traces SF in a more direct way but can be affected by absorption, along with IR emission which is not.
Thus, these hybrid tracers account for the energy lost from the UV/optical light that has suffered absorption by measuring the corresponding IR light.
However, it is relatively harder to have observations in both wavelength regimes due to the lack of deep UV or spectral observations. 
Spectral energy distribution (SED) fitting can provide robust estimations for both the $M_\star$, and SFR, but require extended photometry in multiple bands, while the lack of UV or IR photometry can still lead to large biases \citep[e.g.][]{2013ApJ...768...90L}.

As a result of these limitations, most catalogs providing SFRs, and stellar masses for local Universe galaxies were limited in the SDSS footprint where the sources were mainly characterized through SED fitting combining the optical with WISE \citep{2010AJ....140.1868W} and/or GALEX \citep{2005ApJ...619L...1M} photometry \citep[e.g.][]{2015ApJS..219....8C,2016ApJS..227....2S, 2018ApJ...859...11S}.
Other surveys not limited to the SDSS footprint \citep[e.g.][]{2010PASP..122.1397S, 2011PASP..123.1011A, 2019ApJS..244...24L,2018A&A...620A.112B,2019A&A...624A..80N} combined IR and UV GALEX observations, however, all of them were limited to relatively nearby galaxies and small samples.
Catalogs targeting the whole sky were based solely on IR photometry \citep[e.g.][]{2013AJ....145....6J,2021MNRAS.506.1896K}.

A significant problem in measuring SFRs through monochromatic IR emission is that they do not account for the different degrees of extinction between galaxies, especially in low-metallicity, dust-poor galaxies. 
As a result, an IR SFR indicator would underestimate the SFR for a galaxy with less dust because in this case a larger amount of UV photons, produced by the star-forming activity, would have escaped with respect to a higher-extinction galaxy.
Moreover, the fact that the dust can be excited but other means than reprocessing of UV radiation of young stars \citep[e.g. stochastic heating, heating by older SPs, post-AGB stars, e.t.c.][]{1994ApJ...429..153S,2001ApJ...551..807D,2019A&A...624A..80N,2022arXiv221205688Z} adds another complication for the IR SFR indicators.
In fact, several studies have shown that in quiescent galaxies the mid/far-IR emission is higher with respect to their SFR \citep[e.g.][]{2014MNRAS.444.3427D,2017MNRAS.464.3920S,2019ApJS..244...24L}.
Therefore, for these cases, an IR SFR indicator would overestimate their SFR.

Similarly, the $M_\star$ measurements can be obtained by means of SED fitting, which requires photometry in an extended wavelength range.
However, most commonly they are based on near-IR (NIR) photometry (e.g. 2MASS $K_{\rm S}$, WISE band-1 or band-2) which traces the thermal/continuum emission of low-mass SPs.
Such observations provided $M_\star$ measurements for a large amount of relatively nearby galaxies through the all-sky IR surveys (e.g. IRAS, AKARI, 2MASS, WISE).
However, it has been shown that the relation between NIR emission and the $M_\star$ depends strongly on the age of the SPs.
In the past, optical \citep[e.g.][]{2003ApJS..149..289B,2010RAA....10..329Z,2013MNRAS.433.2946W} or IR colors \citep[e.g.][]{2013AJ....145....6J,2023arXiv230105952J} were used as tracers of the SP age and thus, as a way to correct for their contribution in the NIR emission.
However, these corrections depend on the color of choice and its ability to trace the SP age or the biases induced by interstellar reddening.
The existing calibrations are based on the intrinsic optical colors whereas in their application the observed colors are used.
Therefore, results from various calibrations show discrepancies and large scatter in the estimations of the $M_\star$ \citep[e.g.][]{2021MNRAS.504.3831B}. 
Of course, not accounting for the effect of extinction further exacerbates these discrepancies.

This work investigates the biases introduced by the SP age and extinction in the widely used calibrations of photometric IR luminosity ($L_{\rm IR}$)-to-SFR, and mass-to-light ratio, and proposes new calibrations that try to mitigate these effects.
The analysis is based on creating a large grid of galaxy models with different ISM (e.g. metallicity, ionization parameter U) and dust properties, for a large range of star-forming conditions.
The obtained SEDs and monochromatic luminosities of these galaxy models are then compared with the model SFRs and stellar masses in order to derive the corresponding calibrations.

In Section \ref{sec:biases} we discuss the motivation of this work and give an example that reveals the physical reasons behind the discussed biases.
The used model suite is presented in Section \ref{sec:dataset}.
The details about the analysis and the SFR and $M_\star$ calibrations are provided separately in Section \ref{sec:analysis_results}.
In Section \ref{sec:Observations_comp} the proposed calibrations are compared with previous works and observations, and in Section \ref{sec:conclusions} we present the conclusions.
This work considers only rest-frame luminosities.
Throughout this work, unless stated otherwise, given values correspond to the modes of the distributions and the uncertainties to 68\% percentiles.


\section{Biases in star-formation rate and stellar mass estimations from infrared emission: an example}
\label{sec:biases}

\begin{figure}[ht!]
    \centering
    \includegraphics[width=\columnwidth]{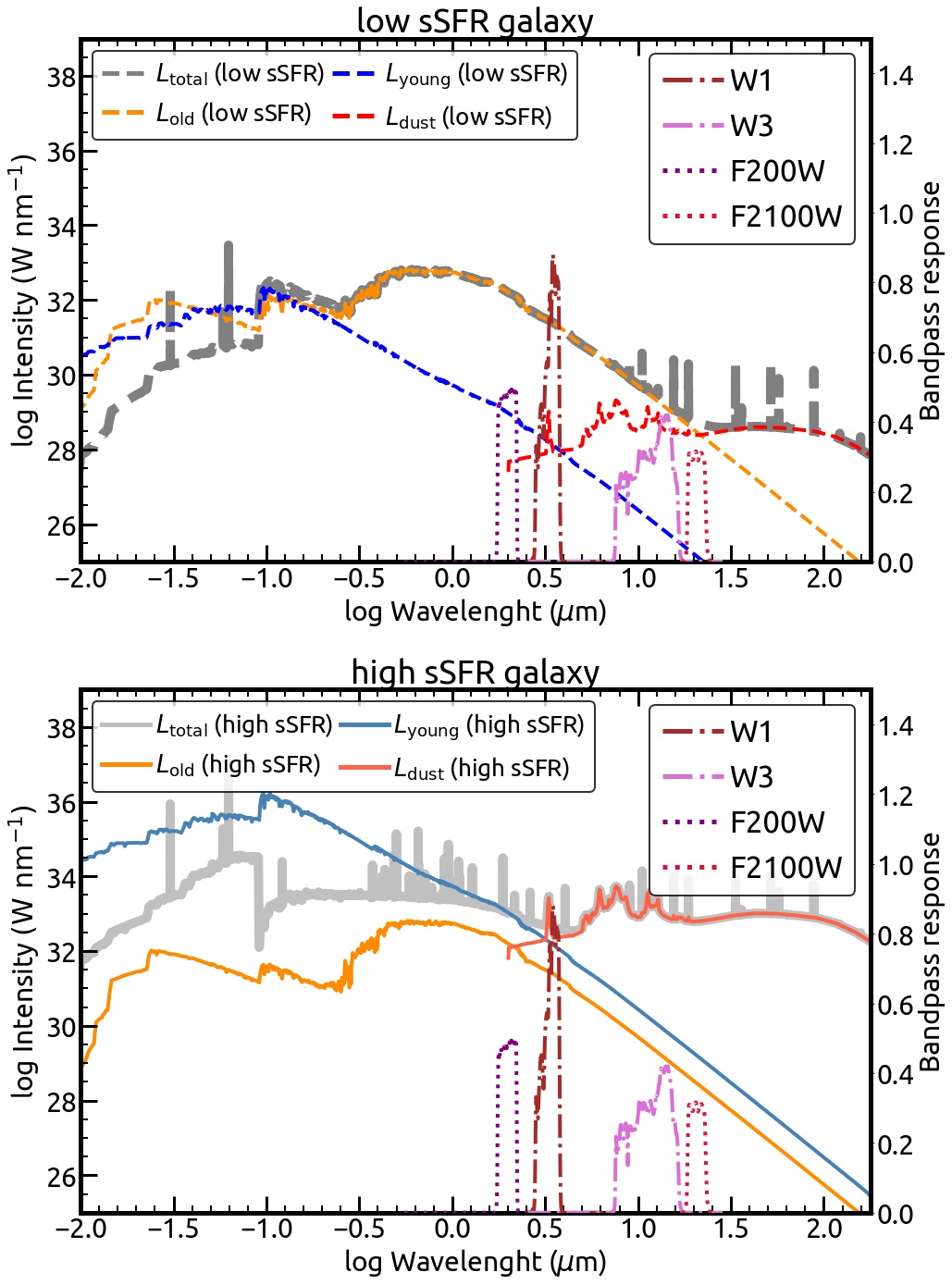}
    \caption{Intensity as a function of wavelength for the total attenuated intensity (gray), and the components of the young stellar (blue), old stellar (orange), and dust emission (red) for two model galaxies: 
    a) in the top panel is a galaxy with low star-forming activity ($\rm SFR = 0.01 ~ M_\odot~yr^{-1}$, $\rm sSFR = 10^{-11.75} ~ M_\odot~yr^{-1}/M_\odot$) and low-extinction ($E(B-V) < 0.001$), and
    b) in the bottom panel is a dusty ($E(B-V) = 0.55$) galaxy with high star-forming activity ($\rm SFR = 68 ~ M_\odot~yr^{-1}$, $\rm sSFR = 10^{-8.5} ~ M_\odot~yr^{-1}/M_\odot$). 
    The dashed-dotted lines with brown and purple colors represent the WISE bands 1 and 3 transmission bandpasses respectively. The dotted fuchsia and red lines represent the JWST NIR-F200W, and MIRI-F2100W bandpasses respectively.}
    \label{fig:SEDs}
\end{figure}

In order to explore the possible biases in measuring the SFR, and $M_\star$ from IR photometry we plot in Figure \ref{fig:SEDs} the young stellar, old stellar, and dust components from the SEDs of two model galaxies with low ($\rm sSFR = 10^{-11.75} ~ M_\odot~yr^{-1}/M_\odot$) and high ($\rm sSFR = 10^{-8.5} ~ M_\odot~yr^{-1}/M_\odot$) specific SFR (${\rm sSFR} \equiv {\rm SFR}/M_\star$) using the code \texttt{CIGALE} \citep{2005MNRAS.360.1413B,2009A&A...507.1793N,2019A&A...622A.103B}. 
Plotted in Figure \ref{fig:SEDs} are also the transmissions of the WISE bands 1 and 3, and JWST NIR-F200W, MIRI-F2100W \citep{2006SSRv..123..485G,2022AJ....163..267G} bandpasses showing the parts of the spectrum used to extract information about the $M_\star$ and the star-forming activity.

These two extreme SEDs demonstrate that the emission in these IR bands corresponds to different physical processes when galaxies are in different star-formation states.
In the NIR (e.g. WISE band-1, JWST NIR-F200W), for the low SF activity galaxy, the stellar emission is orders of magnitude higher compared to the dust emission.
Calibrations of mass-to-light ratio are based on emission in this wavelength range by tracing the thermal emission of the SPs.
However, we see that in a highly star-forming galaxy, the dust component dominates the emission even in the NIR wavelengths, and therefore if it is not taken into account it will significantly bias the stellar-mass estimations.

Similarly for the high SF galaxy, in the mid-IR (MIR; e.g. WISE band-3, JWST MIRI-F2100W) wavelengths, emission from the dust component dominates the SED and it is orders of magnitude higher compared to the stellar continuum emission.
However, we see that for a low-dust/low-sSFR galaxy, the stellar thermal emission can dominate the emission in that bandwidth, enough to significantly bias the SFR estimation.
This example shows that both the SP age and extinction have to be taken into account to properly estimate the SFR and $M_\star$ through the IR emission across the full range of galactic ISM environments.

It should be noted that Figure \ref{fig:SEDs} shows that regardless of the level of the star-forming activity the shape of the spectrum attributed to the dust component is not changing.
This is because \texttt{CIGALE} is not performing radiative transfer that would allow a more detailed description of the SED based on the amount and chemical composition of the dust, the geometric effects, and the local thermal equilibrium of the dust accounting for the SPs in the region of the dust clouds.
However, radiative-transfer analysis is significantly more computationally intensive and could not be performed for this analysis which requires a large number of galaxy models needed to cover the variety of star-formation histories (SFHs) and ISM conditions.
Therefore, in the following analysis dust temperature variations are not being considered, but we expect them to play a minor role in comparison to the total luminosity due to the fact they affect mainly longer wavelengths than 24~$\mu$m \citep[e.g.][]{2021MNRAS.506.3986N}, and the relative scatter in the $L_{\rm IR}$--SFR relation being considerably less compared to that caused by extinction.

\section{Modeling the SED of galaxies}
\label{sec:dataset}

\subsection{Initial library of model-galaxy SEDs}
\label{sec:Init_Lib_SED}

In order to examine the correlation between the IR emission as a tracer of the SFR or the $M_\star$, we created a large library of mock-galaxy SEDs that cover a wide range of SFHs, stellar masses, and ISM conditions.
Modeling galaxies with a SED-fitting code allows us to know a-priori their SFR and $M_\star$, which are an outcome of the given SFHs, while the corresponding band-luminosities are given by the convolution of the filter transmissions and the resulting SEDs.
For the generation of the model-galaxy SEDs, we adopted the commonly used SED fitting code \texttt{CIGALE}.

\texttt{CIGALE} is based on the principle of energy balance, where the energy corresponding to the UV/optical light that has been absorbed by the dust is re-emitted in the mid and far-IR parts of the spectrum.
This makes \texttt{CIGALE} the ideal tool to examine the dependence of IR SFR and $M_\star$ tracers on the amount of dust within a galaxy and the corresponding extinction.
The generated grid of galaxies covered the extinction range for $E(B-V)$ between 0.001 and 1, with steps of $E(B-V)=0.05$ magnitudes.
This $E(B-V)$ refers to the nebular, not the stellar-continuum extinction.

Moreover, a wide range of different SFHs, covering continuously the distribution of star-forming activity between passive and highly star-forming galaxies, is required in order to investigate the dependence of the IR tracers of SFR and $M_\star$ on the SP age.
We adopted the \texttt{sfhdelayed} module, which models the coexistence of an old SP, along with a delayed and recent star-formation burst that generates the young SPs.
By giving a large range to the modulation parameters (Table \ref{tab:pcigale_ini}) the initial library covers a wider range of SFHs and star-forming conditions. 
Throughout this analysis, the SFR coming from modeling with the \texttt{CIGALE} code refers to the \texttt{sfh.sfr10Myrs} that corresponds to the average SFR over 10~Myrs.
In addition, the generated grid of galaxies covered a wide range of metallicities from extremely low ($Z=0.0001$) to high ($Z=0.05$).
The configuration of the SED creation modules is given in Table \ref{tab:pcigale_ini}.
This initial setup generated 1,854,720 SEDs of galaxies.
This library was extended by renormalizing the stellar masses, SFRs, and corresponding luminosities creating a grid of model galaxies with a large range of SFRs, $M_\star$, and ISM conditions.

\begin{figure*}[ht!]
    \centering
    \includegraphics[width=0.7\textwidth]{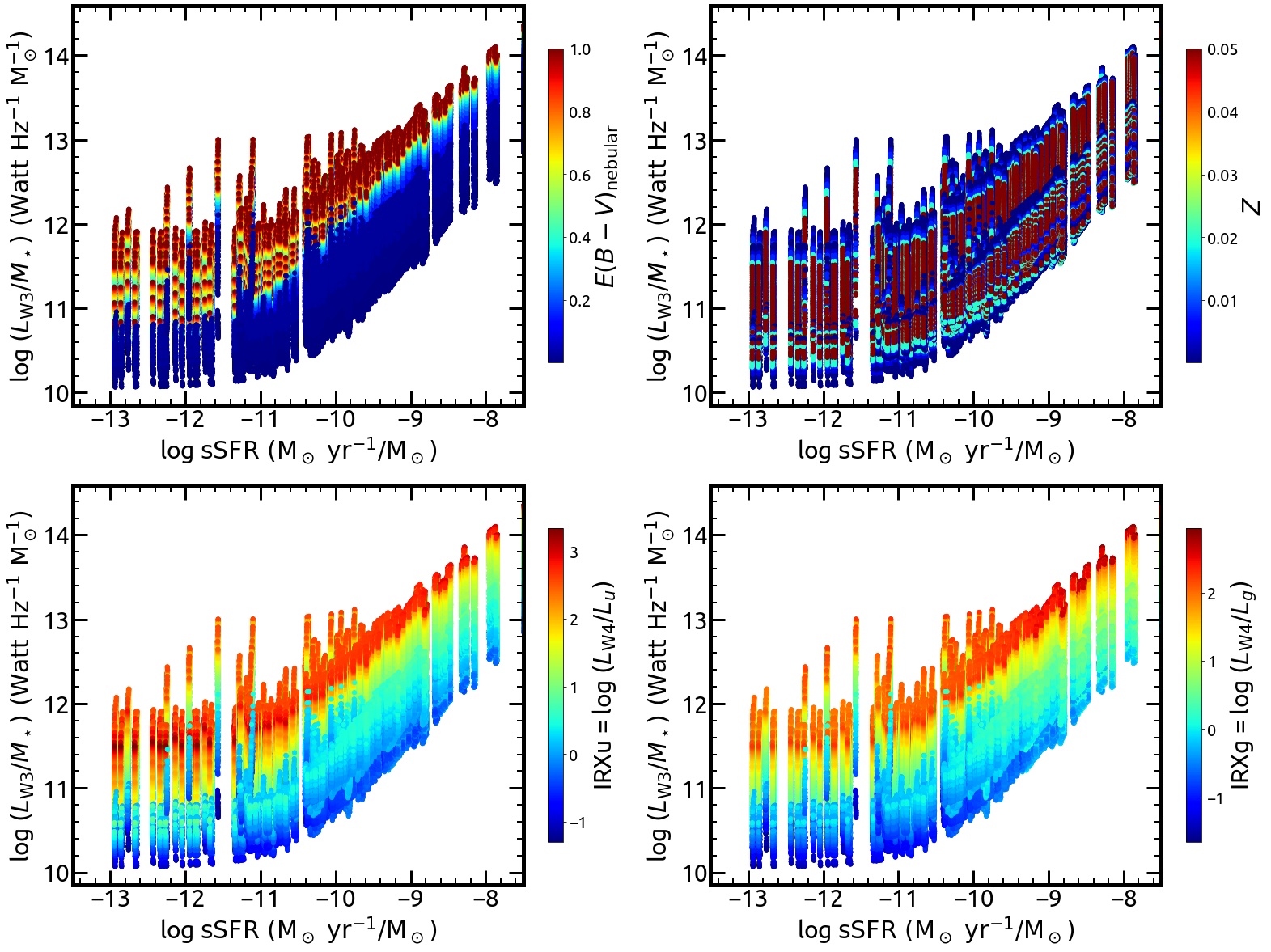}
    \caption{WISE band-3 luminosity ($L_{\rm W3}$) per $M_\star$ unit, as a function of sSFR for the generated library of galaxy SEDs.
    All panels show the same sources but in the top left they are color-coded based on their $E(B-V)$, in the top right based on their metallicity ($Z$), in the bottom left on their infrared excess (${\rm IRXu} = {\rm log}~L_{\rm W4}/L_{u}$), and in the bottom right based on their infrared excess using the g band instead of u (${\rm IRXg} = {\rm log}~L_{\rm W4}/L_{g}$).}
    \label{fig:W3_SFR}
\end{figure*}

Figure \ref{fig:W3_SFR} shows the relation between the luminosity of WISE band-3 ($L_{\rm W3}$) per unit $M_\star$ and the sSFR.
In this work, in addition to the generally used $E(B-V)$ extinction metric that can be obtained through various ways, we also utilize the ratio of WISE band-4 to the optical $u$, or $g$ band luminosities as infrared excess tracers where ${\rm IRXu} = {\rm log}~L_{\rm W4}/L_{u}$ and ${\rm IRXg} = {\rm log}~L_{\rm W4}/L_{g}$.
The latter are chosen because of the large coverage for galaxies in these bands through wide-area surveys \citep[e.g. PANSTARSS;][]{2016arXiv161205560C}.
It should be noted that this IRX is not the same as the one given by CIGALE which refers to the ratio between the dust, and GALEX far-UV (FUV) luminosities.

The generated galaxies show a continuous coverage of sSFRs ranging from extremely low to highly star-forming galaxies.
Moreover, Figure \ref{fig:W3_SFR} shows that the relation between $L_{\rm W3}/M_\star$ and sSFR: a) is not continuously linear: the linearity breaks in low sSFRs where beyond a point the $L_{\rm W3}$ remains constant while the sSFR drops.
Previous works based on observational data \citep[e.g.][]{2016ApJS..227....2S} showed that this break occurs near sSFR~$\simeq 10^{-11} ~ M_\odot~yr^{-1}/M_\odot$ which is similar to what this analysis shows; 
b) there is a strong dependence on extinction which can result in offsets of the sSFR--$L_{\rm W3}/M_\star$ correlation of up to 2 orders of magnitude;
c) metallicity is an additional reason for scatter.
The fact that in low-sSFR galaxies the $L_{\rm W3}$ remains almost constant although the sSFR continues to drop is a representation of the contribution of the old SPs in the MIR emission (see also Section \ref{sec:biases}).
Omitting the common denominator $M_\star$ in the sSFR--$L_{\rm W3}/M_\star$ comparison shows that the above conclusions also stand for the relation between SFR and $L_{\rm W3}$. 

By binning the set of simulated photometries in different metallicity bins and calculating the scatter with respect to the best SFR calibration in each bin we can disentangle the role of metallicity and extinction in the observed scatter. 
We note that this metallicity refers to the metallicity of the stars and because in our simulation the SP and the ISM components are independent there is no intrinsic correlation between the gas-phase metallicity (and therefore the extinction) and the stellar metallicity used as a parameter in our analysis.
The scatter induced by metallicity in the relation between log~SFR and log~$L_{\rm W3}$ is maximized for extremely low SFRs ($\rm SFR \simeq 10^{-6}~M_\odot~yr^{-1}$) and low-extinction ($E(B-V) \simeq 0.1$) galaxies at about 0.3~dex.
Therefore, the metallicity-induced scatter is by far smaller compared to that related to extinction.
In addition, because measuring metallicity requires spectroscopic observations which are more costly, we do not account for the metallicity differences in the following analysis. 
However, we include galaxies that fully cover sub-solar to super-solar metallicities as given by \texttt{CIGALE} ($0.0001 \leq Z \leq 0.05$).
Thus, the following relations and the evaluation of their scatter encompass the dispersion caused by metallicity differences.

\subsection{Fitting sample}
\label{sec:fiting_sample}

The final \textit{master} sample of model-galaxy SEDs used in the following analysis is part of the initial library, and it was selected to fully and evenly cover the plethora of star-forming conditions between passive and starburst, or dwarf and massive galaxies.
It is a result of sampling randomly the initial library in order to produce uniform $M_\star$ and SFR distributions whose ratio, the sSFR, is a distribution that can be described as a normal distribution with mean $\rm \left< log~sSFR/(M_\odot~yr^{-1}/M_\odot) \right> = -10.02$ and $\sigma = 1.45$.
This sample of model-galaxy SEDs covers the range $10^{6.5} < M_\star/(M_\odot) < 10^{12}$, $\rm 10^{-6} < SFR/(M_\odot~yr^{-1}) < 10^{3}$, and $\rm 10^{-13} < sSFR/(M_\odot~yr^{-1}/M_\odot) < 10^{-7.5}$.
Thus, this sample would not bias the analysis towards any particular group of galaxies.
The SFR-$M_\star$ plane for this {\textit{master} sample is shown in Figure \ref{fig:Main_Sequence}.
The best linear fit is:
\begin{eqnarray}
    \rm log~\frac{SFR}{(M_\odot~yr^{-1})} = -8.93~(\pm 0.02) + 0.89 ~(\pm 0.01) ~log~ \frac{M_\star}{(M_\odot)} \quad .
\end{eqnarray}

\begin{figure}[ht!]
    \centering
    \includegraphics[width=0.8\columnwidth]{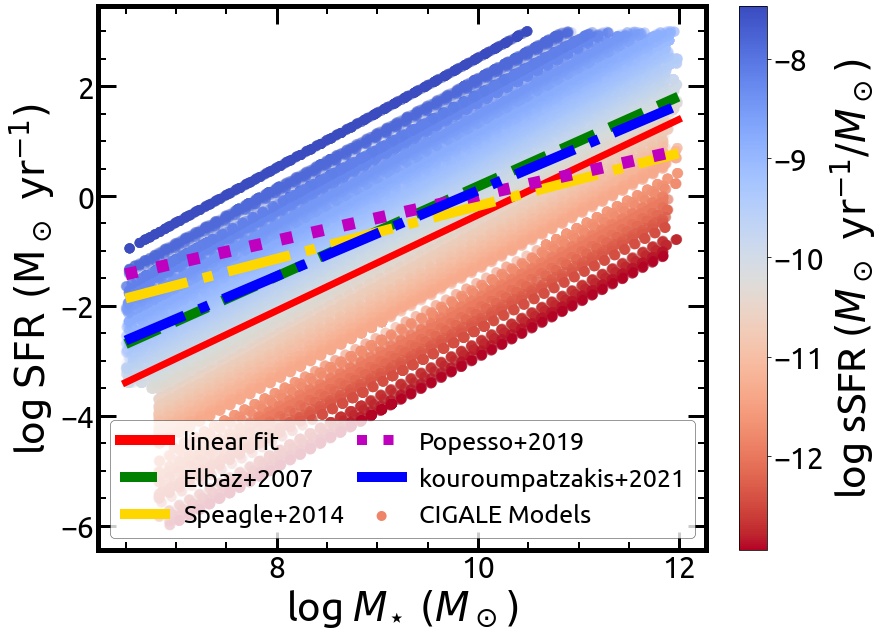}
    \caption{The SFR, $M_\star$ plane for the final sample of \texttt{CIGALE} model galaxies used for the rest of the analysis and fitting. 
    The color code indicates their specific SFR [${\rm sSFR} \equiv {\rm SFR}/M_\star ~ (M_\odot~{\rm yr^{-1}}/M_\odot)$].
    The best linear fit is represented with a red line.
    The green dashed, yellow  dashed-dotted, magenta dotted, and blue dashed-dotted lines represent the main sequence for diverse samples of local Universe galaxies from
    \protect\cite{2007A&A...468...33E}, \protect\cite{2014ApJS..214...15S}, \protect\cite{2019MNRAS.483.3213P}, and \protect\cite{2021MNRAS.506.3079K} respectively.}
    \label{fig:Main_Sequence}
\end{figure}

A goal of this work is to provide an extinction-dependent calibration that can correct for the variance introduced by extinction in the relation between $L_{\rm IR}$ and SFR. However, because estimations of extinction are not always available, this work also provides calibrations that do not necessarily require the input of extinction.
From the \textit{master} sample, two subsets were created in order to account for the differences between the extinction-independent and extinction-dependent calibrations.
For the extinction-dependent relations the sample followed a uniform $E(B{-}V)$ distribution in order for the MCMC fits to be evenly affected by galaxies with low and high extinction.
Thus, the extinction-dependent calibration can compensate and correct for the dependence of the $L_{\rm IR}$--SFR relation on extinction (Figure \ref{fig:W3_SFR}) using as an input the $E(B-V)$ or the IRX.
This \textit{$E(B-V)$-uniform} sample used for the extinction-dependent calibration included 397,256 galaxy models.

The extinction-independent calibration does not have as an input an extinction indicator to correct for the variance it introduces in the $L_{\rm IR}$--SFR relation. 
But still, this calibration is being applied to estimate the SFR of galaxies whatever their extinction may be. 
Thus, the distribution of extinction of the sample used for the fitting of the calibration plays a significant role in determining where the normalization between the $L_{\rm IR}$ and the SFR will be.
The uniform E(B-V) sample has on average more galaxies with higher extinction compared to galaxies found in nature (Fig. \ref{fig:EBV_dist}).
Therefore, if it was used for the fitting, it would have shifted the calibration towards dustier galaxies for whom the SFR corresponds to higher $L_{\rm IR}$ compared to average-extinction galaxies.
This would have mistakenly led to lower normalization, and thus, lower SFRs for the same $L_{\rm IR}$ that would be inappropriate for the bulk of the galaxies. 

Therefore, for the extinction-independent relations, the distribution of the nebular $E(B{-}V)$ of the model galaxies was matched to follow the distribution of $E(B-V)$ of the SDSS spectroscopic \textit{MPA-JHU} catalog \citep{10.1111/j.1365-2966.2003.07154.x,10.1111/j.1365-2966.2004.07881.x,Tremonti_2004}.
The $E(B-V)$ for the SDSS spectroscopic \textit{MPA-JHU} catalog was calculated using the flux ratio of the $\rm H\alpha$ and $\rm H\beta$ emission lines based on the \textit{Balmer decrement} adopting the conversion of \cite{2013ApJ...763..145D}:
\begin{eqnarray}
    E(B{-}V) = 1.97 \, {\rm log} \left[\frac{(f_{\rm H\alpha} / f_{\rm  H\beta} )}{2.86}\right] \quad ,
\end{eqnarray}
using the reddening law of \cite{2000ApJ...533..682C}.
Sources with $E(B-V)/E(B-V)_{\rm err}<3$ were omitted in order to keep only galaxies with reliable extinction estimations.
The distribution of the $E(B-V)$ color excess for the models and the SDSS galaxies (used as a reference) are shown in Figure \ref{fig:EBV_dist}.
This \textit{SDSS-matched} sample included 123,908 galaxy models.

\begin{figure}[ht!]
    \centering
    \includegraphics[width=0.6\columnwidth]{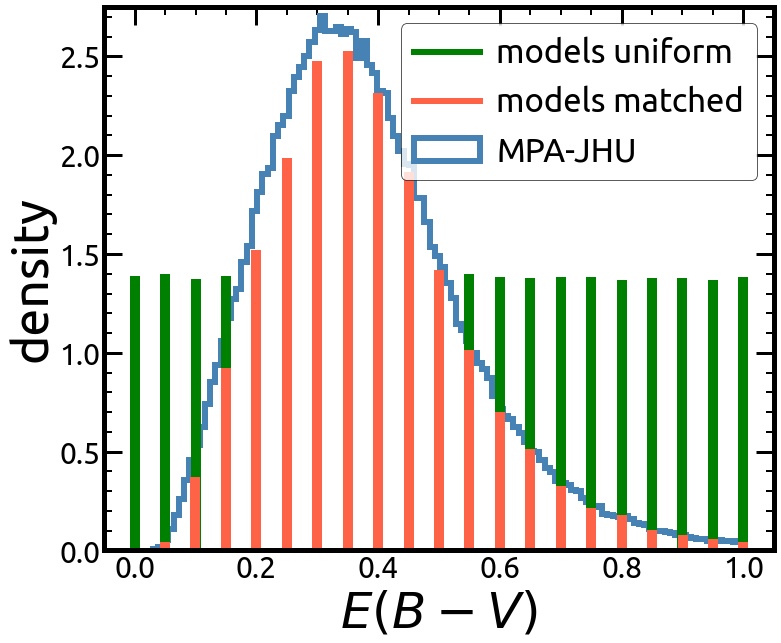}
    \caption{
    The distribution of $E(B-V)$ for the uniform (green histogram) and the SDSS-matched samples (red histogram) of model galaxies.
    With blue color is the nebular $E(B-V)$ distribution based on the Balmer decrement for sources in the SDSS spectroscopic MPA-JHU catalog considering only star-forming galaxies with $E(B-V)/E(B-V)_{\rm err} > 3$.
    The ordinate corresponds to density which is equal to the ratio of the count of each bin to the total number of counts multiplied by the bin width with the integral of the area under the histogram being equal to one.
    }
    \label{fig:EBV_dist}
\end{figure}

Finally, for each of these parent samples, we created by sampling randomly two separate subsets for the fitting and testing of the best-fit models.
The fitting and testing samples included 70\% and 30\% of the parent samples respectively ensuring that there is no overlap between them.

\section{Analysis and results}
\label{sec:analysis_results}

\subsection{Fitting methods}
\label{sec:fiting_methods}

In all cases in the following analyses, the calibration models were fitted with the Markov-chain Monte-Carlo (MCMC) technique using the \texttt{Python emcee} package \citep[][]{emcee}.
The MCMC fitting used 64 walkers over 10000 iterations, while the first 500 were for the burn-in phase.
The initial model parameters were selected by running a maximum likelihood fit on the parameter space based on the \texttt{scipy} minimization algorithm \citep{2020SciPy-NMeth}.
A uniform prior was adopted covering a wide range for each parameter.
The models and best-fit results are presented separately for the $L_{\rm IR}$-to-SFR and the mass-to-light ratios in the following sections.

Additionally, in order to independently test the calibrations given by the MCMC fitting method, we performed machine-learning analysis on the same datasets. 
We adopted the widely-used supervised machine learning algorithm Random Forest (RF) in a regression mode \citep[for a detailed review see][]{RF_review}. 
The building blocks of RF are the decision trees. Each decision tree is a non-parametric model which can be trained to learn the relation between the target variable (e.g SFR, M$_{\star}$) and the feature values (e.g observables as W1, W3, $E(B-V)$, etc.) by using a set of continuous nodes in a tree-like structure. 
During the training, the algorithm searches for the feature and its corresponding value that leads to the best condition of separation at each tree node. 
This process is repeated recursively until a decision tree reaches its final nodes, namely very well-separated and homogenized subsets of the initial training data set. 
We adopted the \textit{Gini impurity} \citep{2019arXiv190407248B} for the calculation of the best separation, which is also the default criterion of the \texttt{sklearn} RF routine.

One of the advantages of the RF regression algorithm is that it can handle problems where the parameter space is highly complex and non-linear, as in our case. 
This makes the RF method the best independent test of the MCMC calibration results which is the standard fitting method followed in this work.

We trained exactly the same calibration datasets that were fitted by MCMC, for consistency between the comparisons of the two methods. 
As with the MCMC method, in all RF models, we considered as training dataset the 70$\%$ of the initial sample and as the test dataset the 30$\%$.
We also tuned the basic hyper-parameters of the RF algorithm (such as the number of trees in the forest, the maximum depth of the tree, etc.) in order to avoid overfitting or underfitting.
We used the implementation of the RF classifier \texttt{sklearn.ensemble.RandomForestClassifier()} provided by the \texttt{scikit-learn}\footnote{\url{https://scikit-learn.org/stable/}} version 1.1.2 \citep{scikit-learn} package for Python 3.

\subsection{Star-formation rate calibrations}
\label{sec:results_SFR}

In order to account for the contribution of old SPs in the MIR, we fit the relation between the MIR luminosity ($L_{\rm MIR}$) and the SFR including a tracer of the $M_\star$.
The NIR (e.g. $K_{\rm S}$, WISE~1 bands) traces part of the stellar continuum emission of low-mass stars and it has been traditionally used to measure the $M_\star$ in galaxies \citep[e.g.][]{2001MNRAS.326..255C,2003ApJS..149..289B}.
In this work, the $L_{\rm MIR}$ and SFR are calibrated including an observable component that traces the $M_\star$. 
This allows to partially disentangle the contribution of old SPs in the MIR emission which can be significant, especially in low-SFR galaxies (see also Section \ref{sec:biases}, Figure \ref{fig:SEDs}).
The form of this relation is described in the following Eq. \ref{eq:IR_SFR_W1_basic}:
\begin{equation}
  {\rm log}L_{\rm MIR} = {\rm log} ( \alpha M_\star + \beta {\rm SFR})  \label{eq:IR_SFR_W1_basic}
\end{equation}
where we assume that the observed luminosity arises from two components: a young component that scales linearly with SFR and an older component that scales linearly with stellar mass.
The older component could be either the tail of the continuum emission of the stars or associated with stochastically heated dust.

As discussed in Section \ref{sec:dataset} our goal is to define a SFR indicator based on NIR and MIR photometry which (a) is applicable to dust rich \textit{and} dust poor galaxies, and (b) galaxies with intense and low-level star-forming activity (where the contribution of an older stellar component in the MIR emission might be not negligible).
Based on the photometric bands we adopt, we have that $M_\star = k~L_{\rm NIR}$. 
We include the effect of dust by parametrizing the parameter $\beta$ in Eq. \ref{eq:IR_SFR_W1_basic} as a logistic function: $\beta = \delta'/(1+\epsilon' ~ 10^{-\zeta' e})$
where $\delta'$, $\epsilon'$, and $\zeta'$ are fitted parameters and $e$ is an extinction metric (Eq. \ref{eq:W3_W1_to_SFR}).
We adopt as an extinction metric the color excess $E(B-V)$ or the IRX index (as defined in Section \ref{sec:dataset}). 
However, because stochastically heated dust may also have a contribution in the NIR bands we also include an extinction term in the parameter $\alpha$.
We find that this is best parametrized with a second-order polynomial of the extinction metric: $\alpha = \alpha' + \beta' e + \gamma' e^2$. 

In order to account for cases where there are no available extinction measurements we also fit the model data with an extinction-independent parametrization, after dropping all the extinction-sensitive terms.
This extinction-independent analysis is performed on the SED models that include the effect of extinction but follow the $E(B-V)$ distribution of SDSS galaxies (Section \ref{sec:fiting_sample}). 

For the WISE photometric system, as a SFR tracer, we adopted the WISE band-3 (12.1~$\mu$m), and as a $M_\star$ tracer the WISE band-1 (3.4~$\mu$m).
The 12.1~$\mu$m band covers a polycyclic aromatic hydrocarbons (PAHs) band which has been used as a SFR indicator \citep[e.g.][]{2013AJ....145....6J}, while the 3.4~$\mu$m band is mostly dominated by stellar continuum emission with a small contribution by a PAH component (e.g. Figure \ref{fig:SEDs}). 
For the JWST photometric system, as a SFR tracer, we adopted the MIRI-F2100W (21~$\mu$m) which probes thermal emission from dust, and as a $M_\star$ tracer the NIR-F200W (2~$\mu$m) which is very similar to the $K_{\rm S}$ band.
Both JWST filters have the best effective responses in the wavelength area of interest, and were designed for general purposes thus, are not significantly dominated by emission from PAHs which could lead to additional biases.

Solving Eq. \ref{eq:IR_SFR_W1_basic} for the SFR term, the calibration of SFR as a function of the measured photometric quantities is:
\begin{equation}
  \begin{aligned}
  \frac{{\rm SFR}}{\rm (M_\odot yr^{-1})} = \frac{L_{\rm MIR} - \alpha L_{\rm NIR}}{10^{21} \beta}\\
    L_{\rm MIR} = \frac{L_{\rm W3}}{{\rm (Watt~Hz^{-1})}},~{\rm or}~L_{\rm MIR} = \frac{L_{\rm F2100W}}{({\rm Watt~Hz^{-1}})}\\
    L_{\rm NIR} = \frac{L_{\rm W1}}{({\rm Watt~Hz^{-1}})},~{\rm or}~L_{\rm NIR} =\frac{L_{\rm F200W}}{{\rm (Watt~Hz^{-1})}}\\
    \alpha = \alpha' + \beta' e + \gamma' e^2\\
    \beta = \frac{\delta'}{1 + \epsilon' 10^{-\zeta' e}}\\
    e = E(B-V),~ e = {\rm IRXu},~ e = {\rm IRXg} \quad .\\ 
  \end{aligned}
  \label{eq:W3_W1_to_SFR}
\end{equation}

The factor $10^{21}$ was introduced in order to homogenize the values of SFR ($\rm M_{\odot}~yr^{-1}$) and the IR luminosities ($\rm Watt~Hz^{-1}$).
Following the method outlined in Section \ref{sec:fiting_methods} we fit the above relation using the \texttt{emcee} implementation of the MCMC fitting scheme. 
Because the photometry is based on SED models given the different SFH scenarios, they come without uncertainties.
Therefore the fitted likelihood functions do not include terms accounting for uncertainties in photometry or in the SFR.
The likelihood function for the extinction-independent model is:
\begin{equation}
    \begin{aligned}
    p(y|x,z,\alpha,\beta) = -\frac{1}{2} \sum_{n} \left[ y_n - \frac{x_n - \alpha z_n}{10^{21} \beta} \right]^2 \quad ,\\
    {\rm and~ for~ the~ extinction-dependent~ model~ is}:\\
    p(y|x,z,e,\alpha',\beta',\gamma',\delta',\epsilon',\zeta') =\\
    -\frac{1}{2} \sum_{n} \left[ y_n - \frac{x_n - (\alpha' + \beta' e + \gamma' e^2) z_n}{10^{21} (\frac{\delta'}{1 + \epsilon' 10^{-\zeta' e}}) } \right]^2 \quad ,\\
    {\rm where: } ~ 
      y = {\rm SFR}/({\rm M_\odot~yr^{-1}})\\
      x = \frac{L_{\rm W3}}{({\rm Watt~Hz^{-1}})} ~ {\rm ,~ or}~ x = \frac{L_{\rm F2100W}} {({\rm Watt~Hz^{-1}})}\\
      z = \frac{L_{\rm W1}}{({\rm Watt~Hz^{-1}})} ~ {\rm ,~ or}~ z =  \frac{L_{\rm F200W}}{({\rm Watt~Hz^{-1}})}\\
      e = E(B-V),~{\rm or}~ e = {\rm IRX} \quad .\\
    \end{aligned}
    \label{eq:likelihood_SFR_with_ext}
\end{equation}
The best-fit results for Eq. \ref{eq:W3_W1_to_SFR} are given in Table \ref{tab:SFR_results}.

\begin{table*}[ht!]
    \renewcommand{\arraystretch}{1.75} 
    \setlength{\tabcolsep}{1.9pt}
    \centering
    \caption{Best-fit results for the Eq. \ref{eq:W3_W1_to_SFR} SFR calibrations.}
    \begin{threeparttable}
    \begin{tabular}{l|cccccc}
        \hline
        \hline
        \multicolumn{7}{c}{WISE band-1, and band-3}\\
        \hline
         & \multicolumn{3}{c}{$\alpha$} & \multicolumn{3}{c}{$\beta$}\\
        $[$W1, W3$]$ & \multicolumn{3}{c}{$0.348 \pm 0.001$}  & \multicolumn{3}{c}{$16.437 \pm 0.002$}\\
        {} & $\alpha'$ & $\beta'$ & $\gamma'$ & $\delta'$ & $\epsilon'$ & $\zeta'$\\
        $[$W1, W3, $E(B{-}V)]$ & $-0.834 \pm 0.000$ &  $2.801 \pm 0.003$ & $2.921 \pm 0.003$ & $16.395 \pm 0.001$ & $6.889 \pm 0.002$ & $9.771 \pm 0.001$\\
        $[$W1, W3, ${\rm IRXu^{\star}}]$ & $0.629 \pm 0.004$ & $-0.625 \pm 0.004$ & $0.509 \pm 0.001$ & $24.112 \pm 0.001$ & $20.695 \pm 0.000$ & $1.068 \pm 0.000$\\
        $[$W1, W3, ${\rm IRXg^{\dag}}]$ & $15.630 \pm 0.004$ & $18.765 \pm 0.004$ & $1.282 \pm 0.001$ & $0.386 \pm 0.001$ & $-0.998 \pm 0.000$ & $0.310 \pm 0.001$\\
        \hline
        \hline
        \multicolumn{7}{c}{JWST NIR-F200W, and MIRI-F2100W}\\
        \hline
          & \multicolumn{3}{c}{$\alpha$} & \multicolumn{3}{c}{$\beta$}\\
        $[$F200W, F2100W$]$ & \multicolumn{3}{c}{$0.215 \pm 0.001$}  & \multicolumn{3}{c}{$27.887 \pm 0.002$}\\
        {} & $\alpha'$ & $\beta'$ & $\gamma'$ & $\delta'$ & $\epsilon'$ & $\zeta'$\\
        $[$F200W, F2100W, $E(B{-}V)]$ & $0.001 \pm 0.000$ &  $-1.944 \pm 0.002$ & $6.135 \pm 0.003$ & $30.511 \pm 0.001$ & $23.391 \pm 0.003$ & $12.798 \pm 0.001$\\
        \hline
        \hline
     \end{tabular}
    \begin{tablenotes}
      \small
      \item ${\rm ^{\star} IRXu} = {\rm log}(L_{\rm W4}/L_{u})$, ${\rm ^{\dag}  IRXg} = {\rm log}(L_{\rm W4}/L_{g})$. 
    \end{tablenotes}
    \end{threeparttable}
    \label{tab:SFR_results}
\end{table*}

\subsubsection{Comparisons with the star-formation rates of the model galaxies}
\label{sec:comps_sfr}

In order to quantify the scatter in the derived SFR calibrations and any biases with respect to the input values or SFRs measured with other methods, we calculate the difference between the results of Eq. \ref{eq:W3_W1_to_SFR} applied to the model SED photometry, and their \textit{true} SFRs which are an outcome of the assumed SFHs.
Table \ref{tab:SFR_fit_comp_res} summarizes the comparisons with the true SFRs for low and high SFRs, and for the uniform, and SDSS-matched extinction samples. 
Comparisons with SFRs calculated using the calibrations of \cite{2013AJ....145....6J}, \cite{2015ApJS..219....8C}, and \cite{2017ApJ...850...68C} are also given as reference.
It should be noted that the 68\% percentiles provided in Table \ref{tab:SFR_fit_comp_res} are dominated by the intrinsic scatter of the data due to the various dependencies that affect the relation between the SFR and the IR luminosities (e.g. extinction distribution of the applied sample, metallicity, ionization parameter, dust content e.t.c.).
On the other hand, 68\% uncertainties given in Table \ref{tab:SFR_results} refer only to the statistical uncertainties of the MCMC fitting procedure and the corresponding parametrization. 

\begin{table*}[ht!]
    \renewcommand{\arraystretch}{1.75} 
    \centering
    \caption{
    The modes and 68\% percentiles for the logarithm of the ratio between the estimated SFR for the model SEDs using Eq. \ref{eq:W3_W1_to_SFR}, based on WISE band-1 and band-3 or NIR-F200W and MIRI-F2100W, over the true SFR as given by the SFHs of the SED modeling. Similarly, estimations based on WISE band-3 for the relations of \cite{2013AJ....145....6J}, \cite{2015ApJS..219....8C}, and \cite{2017ApJ...850...68C} are given. The two columns are for comparisons using the uniform-$E(B-V)$ samples, and the SDSS-matched $E(B-V)$ samples in the SFR range $\rm -4 < log SFR/M_\odot~yr^{-1}) < -1$ (left), and $\rm -1 < log (SFR/M_\odot~yr^{-1})$ (right).
    }
    \begin{tabular}{l|cc|cc}
    \hline
    \hline
    & \multicolumn{4}{c}{$\rm <log \frac{SFR_{model}}{SFR_{true}}>$}\\
   SFR range & \multicolumn{2}{c}{$\rm -4 < log ~ \frac{SFR}{(M_\odot~yr^{-1})}  < -1$} & \multicolumn{2}{c}{$\rm log ~ \frac{SFR}{(M_\odot~yr^{-1})}  > -1$} \\
   \hline
      \diagbox{Model}{$E(B-V)$ distribution}  & uniform & SDSS & uniform & SDSS \\
    \hline
    \hline
    $[$W1, W3] & $ 0.15 ^{+ 1.04 }_{- 0.44 }$ &  $ 0.17 ^{+ 1.11 }_{- 0.2 }$ & $ 0.04 ^{+ 0.36 }_{- 0.38 }$ & $ -0.01 ^{+ 0.32 }_{- 0.16 }$ \\
    $[$W1, W3, $E(B-V)$] & $ 0.07 ^{+ 1.06 }_{- 0.2 }$ & $ 0.05 ^{+ 0.91 }_{- 0.16 }$ & $ -0.07 ^{+ 0.32 }_{- 0.1 }$ & $ -0.01 ^{+ 0.26 }_{- 0.15 }$\\
    $[$W1, W3, ${\rm IRXu}$] & $ 0.03 ^{+ 0.87 }_{- 0.2 }$ & $ 0.04 ^{+ 0.93 }_{- 0.18 }$ & $ -0.08 ^{+ 0.3 }_{- 0.17 }$ & $ -0.08 ^{+ 0.32 }_{- 0.16 }$\\
    $[$W1, W3, ${\rm IRXg}$] & $ 0.16 ^{+ 1.21 }_{- 0.31 }$ & $ -0.01 ^{+ 1.05 }_{- 0.22 }$ & $ -0.17 ^{+ 0.42 }_{- 0.2 }$ & $ -0.32 ^{+ 0.49 }_{- 0.09 }$\\
    $[$F200W, F2100W] &	$ 0.19 ^{+ 1.24 }_{- 0.28 }$ & $ 0.04 ^{+ 1.14 }_{- 0.17 }$ & $ -0.05 ^{+ 0.39 }_{- 0.25 }$ & $ -0.1 ^{+ 0.35 }_{- 0.12 }$\\
    $[$F200W, F2100W, $E(B-V)$] & $0.0 ^{+ 1.17 }_{- 0.18 }$ & $ 0.05 ^{+ 1.23 }_{- 0.19 }$ & $ -0.07 ^{+ 0.29 }_{- 0.15 }$ & $ -0.08 ^{+ 0.33 }_{- 0.17 }$\\
    \cite{2013AJ....145....6J} [W3] & $ 0.04 ^{+ 1.3 }_{- 0.33 }$ & $ -0.06 ^{+ 1.3 }_{- 0.2 }$ & $ -0.3 ^{+ 0.44 }_{- 0.32 }$ & $ -0.23 ^{+ 0.38 }_{- 0.16 }$\\ 
    \cite{2015ApJS..219....8C} [W3] & $ 0.17 ^{+ 1.3 }_{- 0.33 }$ & $ 0.07 ^{+ 1.3 }_{- 0.2 }$ & $ -0.17 ^{+ 0.44 }_{- 0.32 }$ & $ -0.1 ^{+ 0.38 }_{- 0.16 }$\\
    \cite{2017ApJ...850...68C} [W3] & $ 0.76 ^{+ 1.2 }_{- 0.33 }$ & $ 0.63 ^{+ 1.2 }_{- 0.19 }$ & $ 0.3 ^{+ 0.45 }_{- 0.32 }$ & $ 0.26 ^{+ 0.42 }_{- 0.24 }$\\
    \hline
    \hline
    \end{tabular}
    \label{tab:SFR_fit_comp_res}
\end{table*}

Figure \ref{fig:SFR_comp} shows comparisons between the true SFR based on the assumed SFHs ($\rm SFR_{true}$; referring to the average SFR over 10~Myrs from \texttt{CIGALE}), and the calculated SFR based on the model photometry and the application of Eq. \ref{eq:W3_W1_to_SFR} (with the best-fit results; Table \ref{tab:SFR_results}), and the RF models for different extinction values.
As we see in Figure \ref{fig:SFR_comp} and Table \ref{tab:SFR_fit_comp_res} the extinction-dependent relation of Eq. \ref{eq:W3_W1_to_SFR} results in excellent agreement with the true SFR of the galaxies regardless their extinction over the full extinction range. 
Our analysis using the extinction-dependent parametrization provides a remarkable improvement in the reliability of the SFR measurements in low extinction galaxies in comparison to the extinction-independent parametrization or other metrics that do not account for extinction.  
In the case of higher extinction galaxies, the agreement is also excellent (Figure \ref{fig:SFR_comp}, lower-right panel).
In very low SFRs ($\rm \lesssim 10^{-1}~ M_\odot~yr^{-1}$) we see worse performance with a larger scatter.

The extinction-independent relation of Eq. \ref{eq:W3_W1_to_SFR} results in an overall good agreement, that is excellent for galaxies with average, and high extinction.
In very low SFRs the calibrations tend to overestimate the SFRs which, however, cannot be reliably measured also due to SP sampling effects.
This behavior affects all SFR calibrations due to stochasticity driven by relatively small numbers of massive stars in the probed SPs \citep[e.g.][]{2012ARA&A..50..531K}.
The relations based on the JWST photometric system are not shown in Figure \ref{fig:SFR_comp} in order to avoid congestion but they yield similar results as those based on WISE (Table \ref{tab:SFR_fit_comp_res}).

\begin{figure*}
    \centering
    \includegraphics[width=\textwidth]{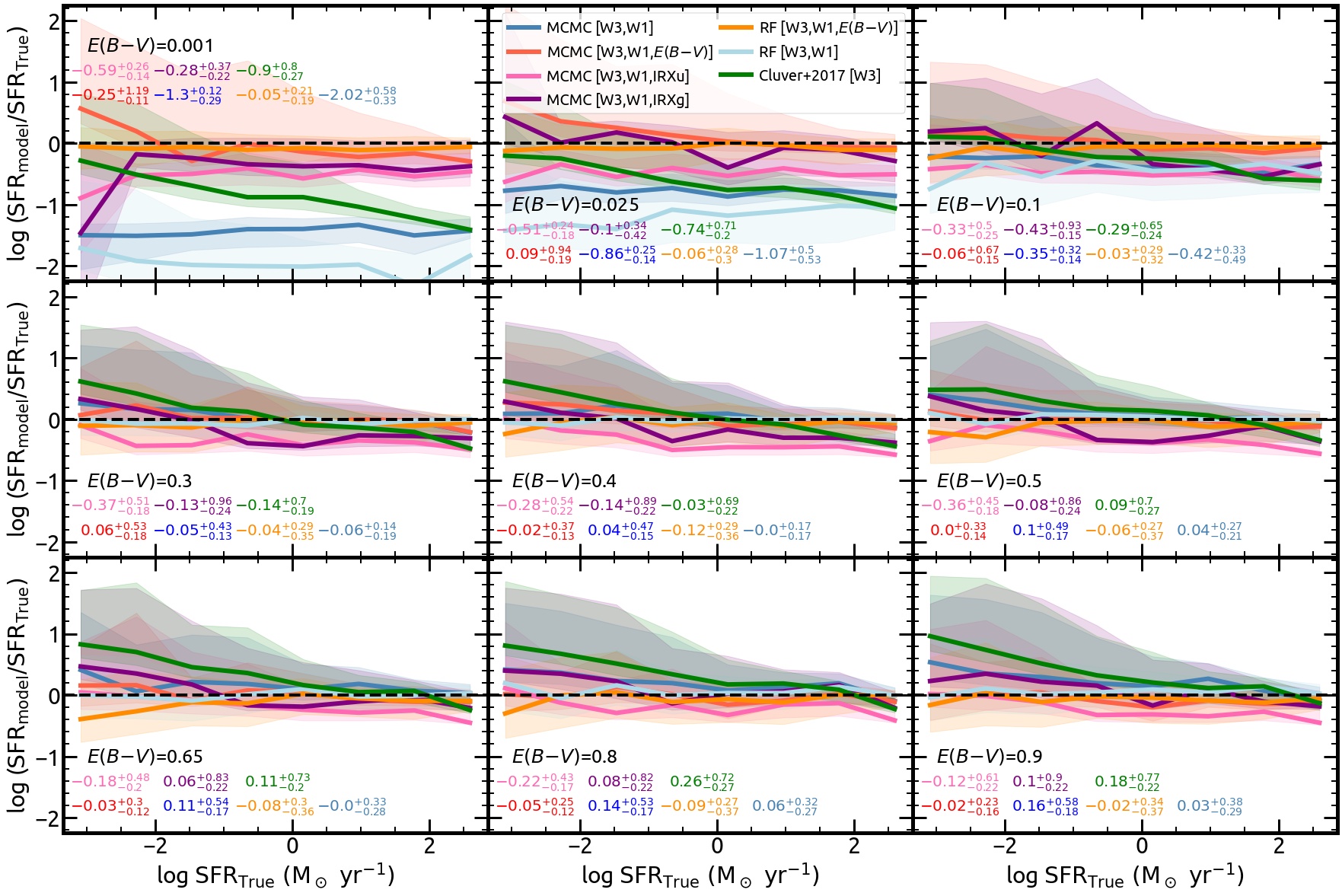}
    \caption{
    Comparison between the true SFR as given by the \texttt{CIGALE} output, and the best-fit models of Eq. \ref{eq:W3_W1_to_SFR} (MCMC; Table \ref{tab:SFR_results}), and the random forest machine learning method (RF) using the WISE bands 1 and 3.
    The $y$-axis represents the logarithm of the ratio between the SFR of each model and the true SFR, thus the black dashed line at zero represents equality.
    Blue, red, pink, and purple colors show the extinction-independent, and extinction-dependent ($E(B-V)$, IRXu, IRXg) relations of Eq. \ref{eq:W3_W1_to_SFR} respectively.
    With light-blue and orange colors are the extinction-independent, and extinction-dependent models based on the RF method respectively.
    The linear log~$L_{\rm W3}$--log~${\rm SFR}$ calibration by \cite{2017ApJ...850...68C} is also displayed in green color.
    The lines represent the distribution modes and the shaded areas show the 68\% percentiles of the $\rm log~ (SFR_{model}/SFR_{true})$ as a function of the log~$\rm SFR_{true}$.
    Each panel shows the comparison of galaxy models with one of the discrete $E(B-V)$ values, shown in the bottom left corner of each panel.
    This comparison is performed with a uniform-$E(B-V)$ sample, therefore, there is an equal number of sources for each extinction value.
    At the bottom of each panel are the modes and 68\% percentiles of the $\rm log~(SFR_{model}/SFR_{True})$ distribution for all galaxy models with the specific $E(B-V)$, shown with corresponding colors to the relations used to derive the SFR.
    These comparisons are also summarized in Table \ref{tab:SFR_fit_comp_res}.
    }
    \label{fig:SFR_comp}
\end{figure*}

The RF method agrees well with the MCMC fitting method with respect to its behavior regarding extinction.
However, for the extinction-dependent calibrations, this method shows slightly improved behavior for model galaxies with extremely low extinction and slightly worse results for model galaxies with extremely high extinction.
Conversely, for extinction-independent calibrations, the RF performs better in high-extinction galaxies and worse in extremely low-extinction galaxies.
This is probably because of the large flexibility of the RF method with respect to MCMC because it does not fit an explicit functional form to the data.
Due to the behavior of the RF models, in the following part of this work we refer to and provide only analytical calibrations which are a result of the MCMC fitting.
However, the fact that the RF method yields similar results to the MCMC indicates that the SFRs from the proposed calibrations of Eq. \ref{eq:W3_W1_to_SFR} do not strongly depend on the fitting method.

The Eq. \ref{eq:W3_W1_to_SFR} relations offer excellent agreement while minimizing the scatter.
It is worth noting that the results between the \textit{true} and inferred SFRs for the extinction dependent and independent relations are almost the same for the SDSS-matched-$E(B-V)$ sample.
This is mainly because the SDSS-matched-$E(B-V)$ distribution mainly consists of average extinction galaxies ($E(B-V) \simeq 0.3$), while it completely lacks galaxies with $E(B-V) < 0.05$ (Figure \ref{fig:EBV_dist}).
Thus, the use of the extinction-independent relation will not significantly bias the SFR estimations for medium extinction galaxies, or samples of galaxies similar to that of SDSS (e.g. local Universe).  
However, for a sample including galaxies with extreme (low or high) values of extinction, the use of the extinction-dependent relations yields better results with less scatter.
As also shown in Figure \ref{fig:SFR_comp} this can be particularly important for dwarf, and dust-free galaxies.

Table \ref{tab:SFR_fit_comp_res} shows the modes and 68\% C.I.s of the distributions of the $\rm log(SFR_{model}/SFR_{true})$ for the calibrations of Eq. \ref{eq:W3_W1_to_SFR}, and that of \cite{2013AJ....145....6J}, \cite{2015ApJS..219....8C}, and \cite{2017ApJ...850...68C} for both the SDSS-matched-$E(B-V)$, and uniform-$E(B-V)$ samples considered here, separated to low, and high-SFR galaxies.
The relation of \cite{2017ApJ...850...68C} shows good agreement for average/high-extinction and high-SFR galaxies but the discrepancy increases as the extinction decreases. 
These comparisons show that the \cite{2017ApJ...850...68C} calibration can lead to an underestimation of the SFR for dust-free galaxies by more than an order of magnitude.
A probable explanation for these discrepancies is the selection of the sample.
The calibration of \cite{2017ApJ...850...68C} used the SINGS/KINGFISH sample of galaxies supplemented with some dwarf and some ultra-luminous IR galaxies (ULIRGs).
On average the agreement is better and the scatter is reduced using the calibrations of Eq. \ref{eq:W3_W1_to_SFR} as shown from the modes and 68\% percentiles of the distributions of $\rm SFR_{model}/SFR_{true}$ (in the bottom of the individual panels; Table \ref{tab:SFR_fit_comp_res}).

Moreover, regardless of the extinction, the commonly used linear $L_{\rm W3}$--SFR calibrations fail to recover the true SFR for passive galaxies.
Figure \ref{fig:SFR_comp} shows that the \cite{2017ApJ...850...68C} calibration overestimates the SFRs in the low-SFR regime regardless of their extinction. 
This is because in low-SFR galaxies (e.g. $\rm \lesssim 0.01 ~ M_\odot~yr^{-1}$) the increasing contribution of the emission from old SPs becomes dominant in the MIR photometric bands like WISE band-3 (e.g. Figures \ref{fig:SEDs}, \ref{fig:W3_SFR}).
On the other hand, the relation of \cite{2017ApJ...850...68C} shows better agreement for the high-SFR sample that follows the SDSS extinction distribution although it still has a tendency to overestimate the SFRs as shown from the upper 68\% percentile of the distribution. 
Overall, the SFRs based on the linear $L_{\rm W3}$--SFR calibrations show a non-linear trend between the estimated SFR and $\rm SFR_{true}$.

As shown in Table \ref{tab:SFR_fit_comp_res} the calibrations of \cite{2013AJ....145....6J}, and \cite{2015ApJS..219....8C} tend to underestimate the SFR on average for high SFR galaxies.
The 68\% percentiles of the ratio between these indicators and the \textit{true} SFR when they are applied to the model SEDs shows that they tend to overpredict the SFR, especially in the case of low-SFR galaxies (Table \ref{tab:SFR_results}). 
This bias is significantly reduced (but not absent) in higher SFR galaxies. 
This behavior is interpreted as the result of older SPs contributing to the emission of the MIR bands we consider.
The calibrations of \cite{2013AJ....145....6J}, and \cite{2015ApJS..219....8C} are not shown in Figure \ref{fig:SFR_comp} in order to avoid congestion.

\subsubsection{Mid-IR emission dominated by old stellar populations}
\label{sec:MIR_old_young_dependence}

Although the data used to derive the parameters of  Eq. \ref{eq:W3_W1_to_SFR} are based on SED fitting tools built on the principle of energy balance, the large scatter in the W3 luminosity at low SFR (SFR~$\rm <0.01~M_\odot~yr^{-1}$) combined with the negative term in Eq. \ref{eq:W3_W1_to_SFR} (introduced to account for the contribution of old SPs) may lead to negative SFR for galaxies with low SFR or low sSFR. 
This is because Eq. \ref{eq:W3_W1_to_SFR} gives an average relation between SFR, MIR, and NIR luminosity while the significant scatter resulting from the wide variety of SFHs and ISM parameters may result in stronger NIR emission with respect to the MIR emission. 
This effect may be exacerbated by the presence of stochastically heated dust (which is more likely in low sSFR galaxies). 
The 2--4 $\mu$m bands include a strong dust emission feature that can be significantly excited by the general interstellar radiation field \citep[e.g.][]{2015A&A...580A..87C}.
For these cases, the SFR can not be calculated through Eq. \ref{eq:W3_W1_to_SFR}.

The comparisons with the test dataset (Section \ref{sec:fiting_methods}) show that sources showing negative SFRs using Eq. \ref{eq:W3_W1_to_SFR} have on average $L_{\rm old} \simeq 65 ~ L_{\rm young}$.
The results are similar for both the extinction-dependent and extinction-independent relations.
For the extinction-independent calibration the cutoff value for the W3--W1 color, below which galaxies give negative SFRs, is ${\rm log} (L_{\rm W3}/L_{\rm W1}) = -0.371$, while this is not constant for the extinction-dependent relations.
The galaxies whose SFR can not be measured are relatively few (${\sim} 6 \%$, and  ${\sim} 12 \%$ for the extinction-independent and extinction-dependent relations respectively) considering that this analysis included galaxies with extremely low sSFR down to sSFR~$\rm \simeq 10^{-13}~M_\odot~yr^{-1}/M_\odot$.
Therefore, galaxies that yield negative SFRs based on Eq. \ref{eq:W3_W1_to_SFR} have extremely small star-forming activity and their IR emission is dominated by old SPs.
Overall, the Eq. \ref{eq:W3_W1_to_SFR} calibrations offer reliable estimations even in low SFRs.

Additionally, we examine the capability of the IR--SFR calibrations to recover the SFR throughout the range of SF activity.
Figure \ref{fig:SFR_to_sSFR_model_com} shows the ratio of the estimated SFR, from the proposed and existing IR--SFR calibrations, and the true SFR as a function of the model-galaxies sSFR. 
Previous works already showed that in passive galaxies with $\rm sSFR \lesssim 10^{-11} ~ M_\odot~yr^{-1}/M_\odot$ the dust is not heated by young star and thus, SFR estimations based on the MIR emission for these galaxies are not reliable \citep[e.g.][]{1998ARA&A..36..189K,2009ApJ...700..161S, 2016ApJS..227....2S}. 
Figure \ref{fig:SFR_to_sSFR_model_com} shows that all the IR--SFR calibrations compared here fail to recover and overestimate the SFR in galaxies with $\rm sSFR \lesssim 10^{-10.7} ~ M_\odot~yr^{-1}/M_\odot$.
However, the proposed calibrations from this work, and especially the extinction dependent, tend to reduce the biases in the SFR estimation.
This is demonstrated in Figure \ref{fig:SFR_to_sSFR_model_com} by the slightly better agreement between the modes of the proposed calibrations (red and blue lines) with the equality (black dashed) line in relatively low sSFRs ($\rm sSFR \simeq 10^{-10.5} ~ M_\odot~yr^{-1}/M_\odot$) and in high sSFRs ($\rm sSFR \simeq 10^{-8} ~ M_\odot~yr^{-1}/M_\odot$).

\begin{figure*}[ht!]
    \centering
    \includegraphics[width=\textwidth]{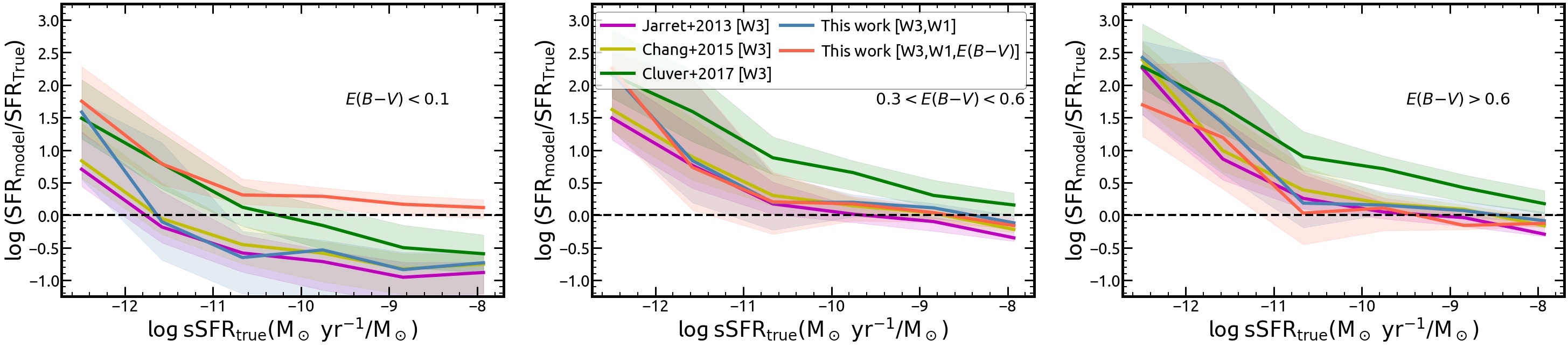}
    \caption{
    The logarithm of the ratio of the estimated SFR based on the IR emission over the true SFR as given by the \texttt{CIGALE} models, as a function of their true specific SFR (sSFR).
    The comparisons show the extinction-dependent and extinction-independent calibrations of Eq. \ref{eq:W3_W1_to_SFR}, and the $L_{\rm IR}$--SFR calibrations of \cite{2013AJ....145....6J}, \cite{2015ApJS..219....8C}, and \cite{2017ApJ...850...68C} with blue, red, purple, yellow, and green lines respectively for model galaxies separated in groups of low (left), moderate (center), and high (right) extinction.}
    \label{fig:SFR_to_sSFR_model_com}
\end{figure*}

\subsection{Stellar mass calibrations}
\label{sec:results_stellar_mass}

Similarly to the SFR-$L_{\rm IR}$ relation, we explore the dependence of the mass-to-light ratio on the SP age and dust content/extinction.
For this analysis, we use the same data as with the SFR analysis, as described in Section \ref{sec:fiting_sample}.
Previous studies have identified the SP age as the main cause of uncertainty in the calculation of the $M_\star$ \citep[e.g.][]{1993ApJ...418..123R}.
Here, as an observational tracer of the SP age, we adopt the $u{-}r$, and $g{-}r$ colors, which have also been used in other studies \citep[e.g.][]{2003ApJS..149..289B,2013MNRAS.433.2946W}.
However, this work also explores the possibility of increased scatter in the inferred mass-to-light ratio induced by the use of extinction-corrected colors as is usually the case.
Dust emission, heated by the star-forming activity is mainly in the mid, and far-IR parts of the spectrum.

However, emission from PAH molecules and stochastically heated dust grains may have a significant contribution to the NIR emission. 
The relation between the mass-to-light ratio with the optical colors shows a large scatter (e.g. Figure \ref{fig:ML_EBV}).
This may be the result of extinction affecting the $u{-}r$, and $g{-}r$ colors of the SPs creating multiple mass-to-light ratio tracks for different values of extinction.
In fact, if we account for the extinction (i.e. correct the $u{-}r$ color for the extinction based on the $E(B-V)$) the scatter is slightly reduced. 
The advantage of our analysis is that it accounts for the effect of dust in two ways: (a) by its effect on the colors of the SPs and (b) by its effect on the NIR emission through stochastically heated dust.
There is a tertiary effect originating from the contribution of the H$\alpha$ emission in the $r$ band photometry. 
This is also indirectly accounted for through the (loose) correlation between extinction in the ISM and star-forming activity.

Figure \ref{fig:ML_EBV} shows the mass-to-light ratio of the \texttt{CIGALE} model galaxies, as a function of the $u{-}r$ color.
Each point represents a single model SED and it is color-coded based on its extinction.
Figure \ref{fig:ML_EBV} reveals the scatter induced by the differences in extinction, where especially for blue galaxies it can be more than an order of magnitude.
The scatter increases for decreasing $u{-}r$ color.
However, in highly star-forming galaxies ($u-r \lesssim 1.5$ or $g-r \lesssim 0.75$) the scatter is by far larger.
In these color regimes, galaxies have higher SFRs, and therefore, IR emission from the dust is brighter, and its respective dependence on extinction is stronger.
Moreover, the $g{-}r$ color has an overall narrower range and shows a larger scatter as a function of the mass-to-light ratio compared to $u{-}r$, and therefore is not as good as a SP age indicator for the purposes of this work.

\begin{figure*}
    \centering
    \begin{tabular}{@{}c@{}}
    \includegraphics[width=\columnwidth]{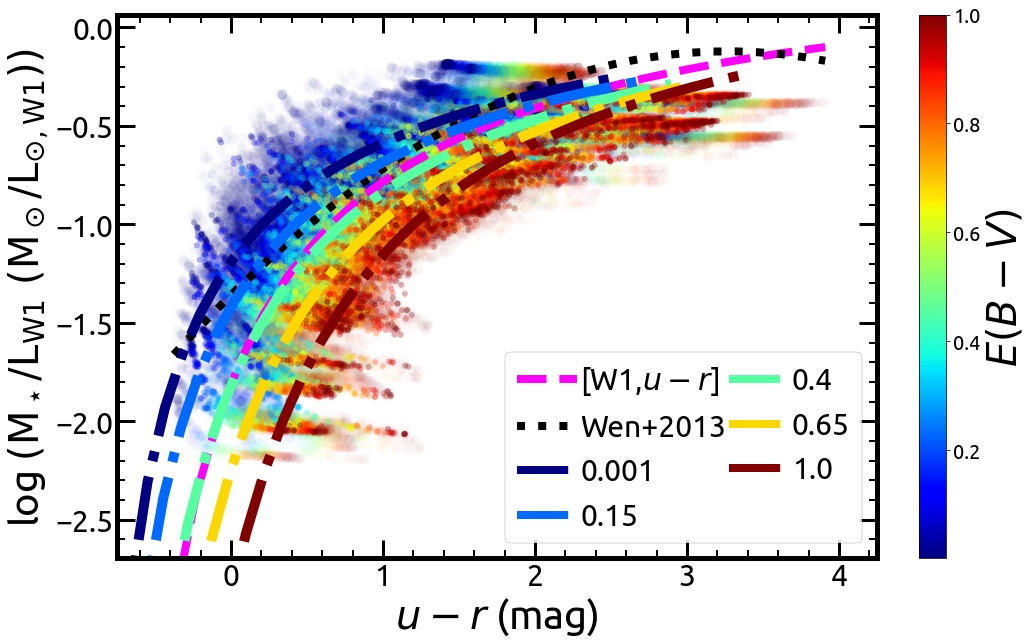}
    \includegraphics[width=\columnwidth]{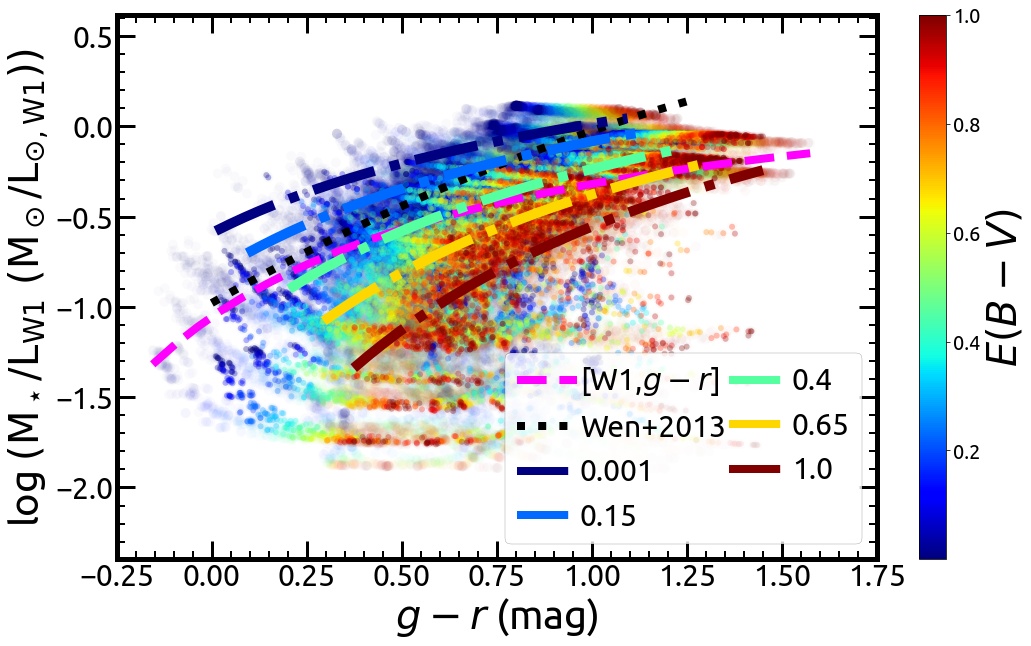}
    \end{tabular}
    \begin{tabular}{@{}c@{}}
    \includegraphics[width=\columnwidth]{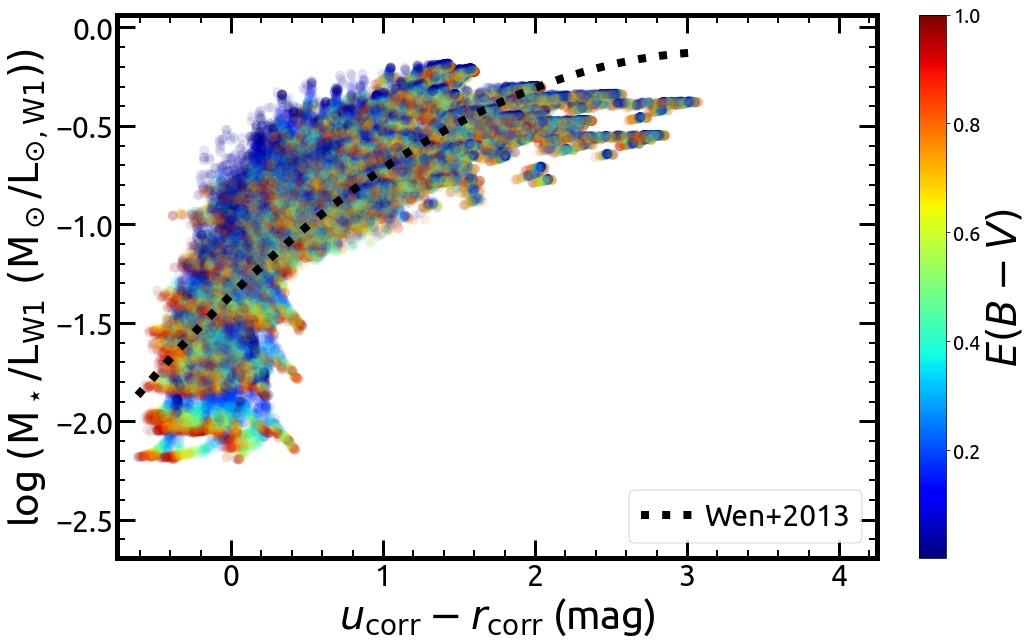}
    \includegraphics[width=\columnwidth]{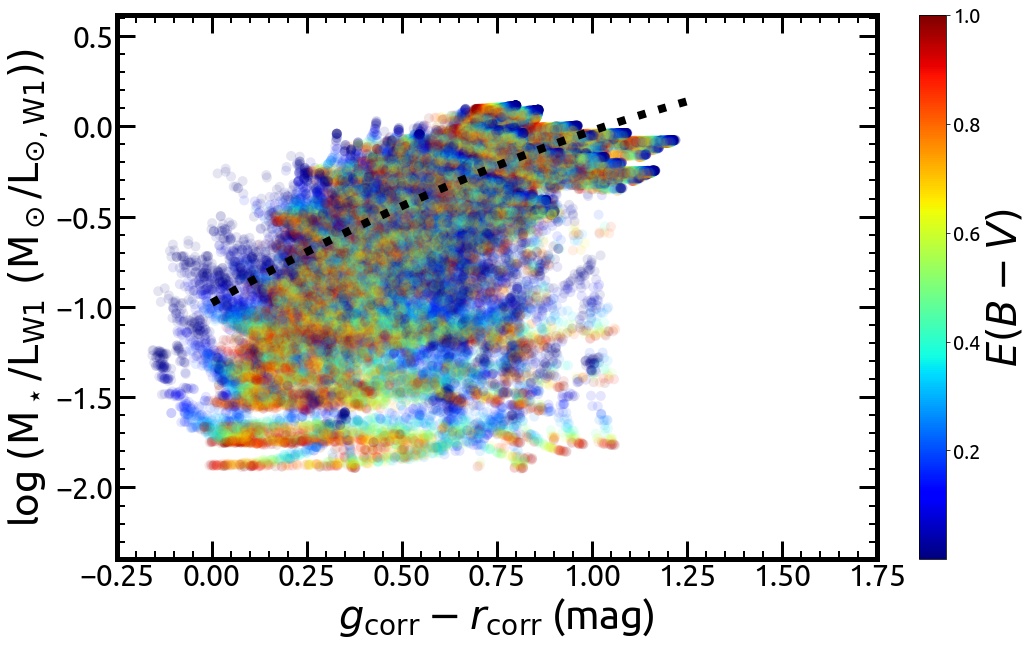}
    \end{tabular}
    \caption{Mass-to-light ratio as a function of $u{-}r$ (left column) and $g{-}r$ (right column) color for the SDSS-matched sample of the modeled galaxies.
    In the bottom row, the $u{-}r$ and $g{-}r$ colors are corrected for extinction while in the top row, they are not.
    The galaxies are color-coded based on their $E(B-V)$.
    The dashed magenta line represents the extinction-independent relation of Eq. \ref{eq:ML_ratio}, and the dashed-dotted lines the extinction-dependent, drawn for some discrete $E(B-V)$ values as shown in the legend using the parameters presented in Table \ref{tab:Mstar_results}.
    The dotted black line shows the relations of \cite{2013MNRAS.433.2946W}.
    }
    \label{fig:ML_EBV}
\end{figure*}

We describe the correlation between the mass-to-light ratio and the optical color as a power law with negative power values, thus as $M/L = \alpha (u-r)^\beta$.
Moreover, in this analysis, we fit the mass-to-light ratio relation with the optical color taking also into account the extinction.
We include the effect of dust by parametrizing the parameter $\beta$ as a linear function of extinction with the form $\beta = \delta' + \epsilon' ~ e$ where $\delta'$ and $\epsilon'$ are fitted parameters and $e$ is an extinction metric.
We adopt as an extinction metric ($e$) the color excess $E(B-V)$ or the IRX index (as defined in Section \ref{sec:dataset}). 
However, because extinction also affects the $u{-}r$ color we also include an extinction term in the parameter $\alpha$ which is best parametrized with a second-order polynomial of the extinction metric: $\alpha = \alpha' + \beta' e + \gamma' e^2$. 
In order to account for cases where there are no available extinction measurements we also fit the data with an extinction-independent parametrization.
These fits involve the SED models that include the effect of extinction but follow the $E(B-V)$ distribution of SDSS galaxies (Section \ref{sec:fiting_sample}).

The final mass-to-light calibrations are presented in Eq. \ref{eq:ML_ratio}:
\begin{equation}
  \begin{aligned}
  {\rm log}~\frac{M_\star/{(\rm M_\odot)}}{L_{\rm NIR}} = 
  \alpha (x + 1)^{\beta} + 0.5\\
  x = u-r, ~{\rm or}~ x = g-r \\
  L_{\rm NIR} = \frac{L_{\rm W1}}{(L_{\rm \odot, W1})}, ~{\rm or}~ L_{\rm NIR} =  \frac{L_{\rm F200W}}{(L_{\rm \odot})}\\
  \alpha = \alpha' + \beta' e + \gamma' e^2\\
  \beta = \delta' + \epsilon' e\\
  e = E(B-V), ~{\rm or}~ e = {\rm IRXu}, ~{\rm or}~ e = {\rm IRXg}
  \end{aligned}
  \label{eq:ML_ratio}
\end{equation}
The addition of 1 to the $u{-}r$ or $g{-}r$ color, and the term $0.5$ were introduced in order to avoid singularities in the model.
$L_{\rm W1}$ is measured in the in-band equivalent solar luminosity \citep[see e.g.][]{2013AJ....145....6J}.
It is equal to scaling $\nu L_{\nu}$ by 22.883.
The $L_{\rm F200W}$ is measured in solar luminosities.
The best-fit results based on the MCMC fitting method for the parameters $\alpha, ~ \beta$ and $\alpha',~ \beta', ~ \gamma',~ \delta',~ \epsilon'$ are given in Table \ref{tab:Mstar_results}.
Moreover, we performed a RF model for the same datasets and the same sets of features with the MCMC in order to independently compare the result of the MCMC method.

\begin{table*}[ht!]
    \renewcommand{\arraystretch}{1.5} 
    \centering
    \caption{Best-fit results for the Eq. \ref{eq:ML_ratio} mass-to-light ratio calibrations.}
    \begin{threeparttable}
    \begin{tabular}{l|ccccc}
        \hline
        \hline
        Model & \multicolumn{5}{c}{WISE band-1}\\
        \hline
          & \multicolumn{3}{c}{$\alpha$} & \multicolumn{2}{c}{$\beta$}\\
         $[$W1, $u{-}r$] & \multicolumn{3}{c}{$-2.307\pm 0.013$}  & \multicolumn{2}{c}{$-0.847 \pm 0.010$}\\
        {} & $\alpha'$ & $\beta'$ & $\gamma'$ & $\delta'$ & $\epsilon'$ \\
        $[$W1, $u{-}r$, $E(B-V)]$ & $-1.689^{+0.010}_{-0.011}$ & $-1.493^{+0.048}_{-0.047}$ & $-0.201^{+0.061}_{-0.058}$ & $-0.637^{+0.009}_{-0.009}$ & $-0.394^{+0.016}_{-0.017}$\\
        $[$W1, $u{-}r$, $\rm IRXu^{\star} ]$ & $-1.520 \pm 0.013$ & $-0.201 \pm 0.010$ & $-0.168 \pm 0.05$ & $-0.532 \pm 0.010$ & $-0.133 \pm 0.006$ \\
        \hline
          & \multicolumn{3}{c}{$\alpha$} & \multicolumn{2}{c}{$\beta$}\\
        $[$W1, $g{-}r$] & \multicolumn{3}{c}{$-1.555^{+0.005}_{-0.006}$}  & \multicolumn{2}{c}{$-0.924^{+0.013}_{-0.010}$}\\
        {} & $\alpha'$ & $\beta'$ & $\gamma'$ & $\delta'$ & $\epsilon'$\\
        $[$W1, $g{-}r$, $E(B-V)]$ & $-1.266^{+0.006}_{-0.007}$ & $-0.765^{+0.034}_{-0.034}$ & $-0.115^{+0.037}_{-0.036}$ & $-0.608^{+0.012}_{-0.012}$ & $-0.896^{+0.024}_{-0.023}$\\
        $[$W1, $g{-}r$, $\rm IRXg^{\dag} ]$ & $-1.135^{+0.005}_{-0.005}$ &  $-0.135^{+0.006}_{-0.005}$ & $-0.156^{+0.003}_{-0.003}$ & $-0.661^{+0.013}_{-0.013}$ & $-0.106^{+0.009}_{-0.009}$ \\
        \hline
        \hline
        & \multicolumn{5}{c}{JWST NIR-F200W}\\
        \hline
         & \multicolumn{3}{c}{$\alpha$} & \multicolumn{2}{c}{$\beta$}\\
         $[$F200W, $u-r]$ & \multicolumn{3}{c}{$-1.193\pm 0.015$}  & \multicolumn{2}{c}{$-1.691 \pm 0.028$}\\
        {} & $\alpha'$ & $\beta'$ & $\gamma'$ & $\delta'$ & $\epsilon'$\\
         $[$F200W, $u{-}r$, $E(B-V)]$ & $-0.838^{+0.012}_{-0.012}$ & $-0.800^{+0.079}_{-0.078}$ & $-0.815^{+0.114}_{-0.114}$ & $-1.509^{+0.029}_{-0.030}$ & $-0.719^{+0.059}_{-0.061}$\\
         \hline
         & \multicolumn{3}{c}{$\alpha$} & \multicolumn{2}{c}{$\beta$}\\
         $[$F200W, $g-r]$ & \multicolumn{3}{c}{$-0.547\pm 0.005$}  & \multicolumn{2}{c}{$-1.843 \pm 0.031$}\\
        {} & $\alpha'$ & $\beta'$ & $\gamma'$ & $\delta'$ & $\epsilon'$\\
         $[$F200W, $g{-}r$, $E(B-V)]$ & $-0.455^{+0.007}_{-0.007}$ & $-0.148^{+0.038}_{-0.038}$ & $-0.376^{+0.042}_{-0.042}$ & $-1.123^{+0.030}_{-0.029}$ & $-2.542^{+0.071}_{-0.069}$\\
        \hline
        \hline
     \end{tabular}
     \begin{tablenotes}
      \small
      \item ${\rm ^{\star} IRXu} = {\rm log}(L_{\rm W4}/L_{u})$, ${\rm ^{\dag}  IRXg} = {\rm log}(L_{\rm W4}/L_{g})$. 
    \end{tablenotes}
    \end{threeparttable}
    \label{tab:Mstar_results}
\end{table*}

As mentioned in Section \ref{sec:results_SFR}, because the SEDs are an output of modeling, their photometries, and their stellar masses come without uncertainties, and thus they are not included in the likelihoods of the MCMC fitting.
The likelihood function for the extinction-independent model is:
\begin{equation}
    \begin{aligned}
    {\rm log}~ p(y|x,\alpha,\beta) = -\frac{1}{2} \sum_{n} \left( y_n - \alpha x_n^{\beta} \right)^2\\
    {\rm and~ for~ the~ extinction{-}dependent~ model~ is:}\\
    {\rm log}~ p(y|x,e,\alpha',\beta',\gamma',\delta',\epsilon') =\\
    -\frac{1}{2} \sum_{n} \left[ y_n - 
   (\alpha' + \beta' e + \gamma' e^{2})~x_n^{(\delta'+\epsilon' e)}\right]^2\\
   {\rm where:} \\
   y = {\rm log}~\frac{M_\star/L_{\rm W1}}{(M_\odot/L_{\rm \odot, W1})} - 0.5, ~{\rm or}~ y ={\rm log}~\frac{M_\star/L_{\rm F200W}}{(M_\odot/L_{\rm \odot})} - 0.5\\
   x = u-r+1~({\rm mag}), ~{\rm or}~ x = g-r+1~({\rm mag}) \quad .
    \end{aligned}
    \label{eq:mstar_likelihood}
\end{equation}

\subsubsection{Comparisons with the stellar masses of the model galaxies}
\label{sec:comps_mstar_model_gals}

Figure \ref{fig:ML_EBV} shows the extinction-independent relation of  Eq. \ref{eq:ML_ratio} with a dashed magenta line.
The dashed-doted lines show the extinction-dependent relation for some representative extinction values.
The extinction-independent relation follows the median position of the points.
The lines drawn based on the extinction-dependent relation follow well the data points of similar extinction values.
As a reference, we also compare with the calibration of \cite{2013MNRAS.433.2946W} which is empirically calibrated based on the W1 emission and stellar masses estimated from the \textit{MPA–JHU} analysis.
The mass-to-light ratio in this relation is parametrized as a polynomial function of the $u{-}r$ or $g{-}r$ colors.
The relation of \cite{2013MNRAS.433.2946W} shows good agreement for average $u{-}r$ colors ($u-r \simeq 1.5$), as well as, for moderate extinction [$E(B-V) \simeq 0.5$].
However, it fails to recover the $M_\star$ for galaxies with low or high extinction, while it shows larger discrepancies near the lower and upper limits of $u{-}r$ color.

Figure \ref{fig:ML_ratio} shows the logarithm of the ratio between the calculated stellar masses for the MCMC, and RF models (using the WISE band-1) over the true $M_\star$, as a function of the true $M_\star$.
The true $M_\star$ is an output of \texttt{CIGALE} based on the given SFHs of the modeled galaxies.
We also plot the calculated stellar masses based on the relation of \cite{2013MNRAS.433.2946W} as a comparison reference.
Figure \ref{fig:ML_ratio} does not show the results based on the JWST-NIR F200W band for clarity because these are very similar to those using the WISE band-1.
Table \ref{tab:Mstar_dist_results} gives the comparisons for the proposed relations for all the models considered in this analysis, separated in low $u{-}r$ (younger SPs or lower extinction), and high $u{-}r$ (older SPs or higher extinction) color.
These comparisons are also given separately for the sample having a uniform distribution in extinction, and the SDSS-matched sample (Section \ref{sec:fiting_sample}) in order to reveal possible biases based on the selected sample of galaxies.

The MCMC relations show robust estimations throughout the $M_\star$ range. 
Figure \ref{fig:ML_ratio} demonstrates that the use of the extinction feature in the $M_\star$ calculations is important in low and high-extinction galaxies. 
The extinction-independent relation of Eq. \ref{eq:ML_ratio} is in good agreement for mid-range and high-extinction sources.
However, in all cases, the scatter is increased when the extinction is not taken into account.
Overall, the RF yields similar results to the MCMC indicating the results of the proposed calibrations are independent of the method.

\begin{figure*}
    \centering
    \includegraphics[width=\textwidth]{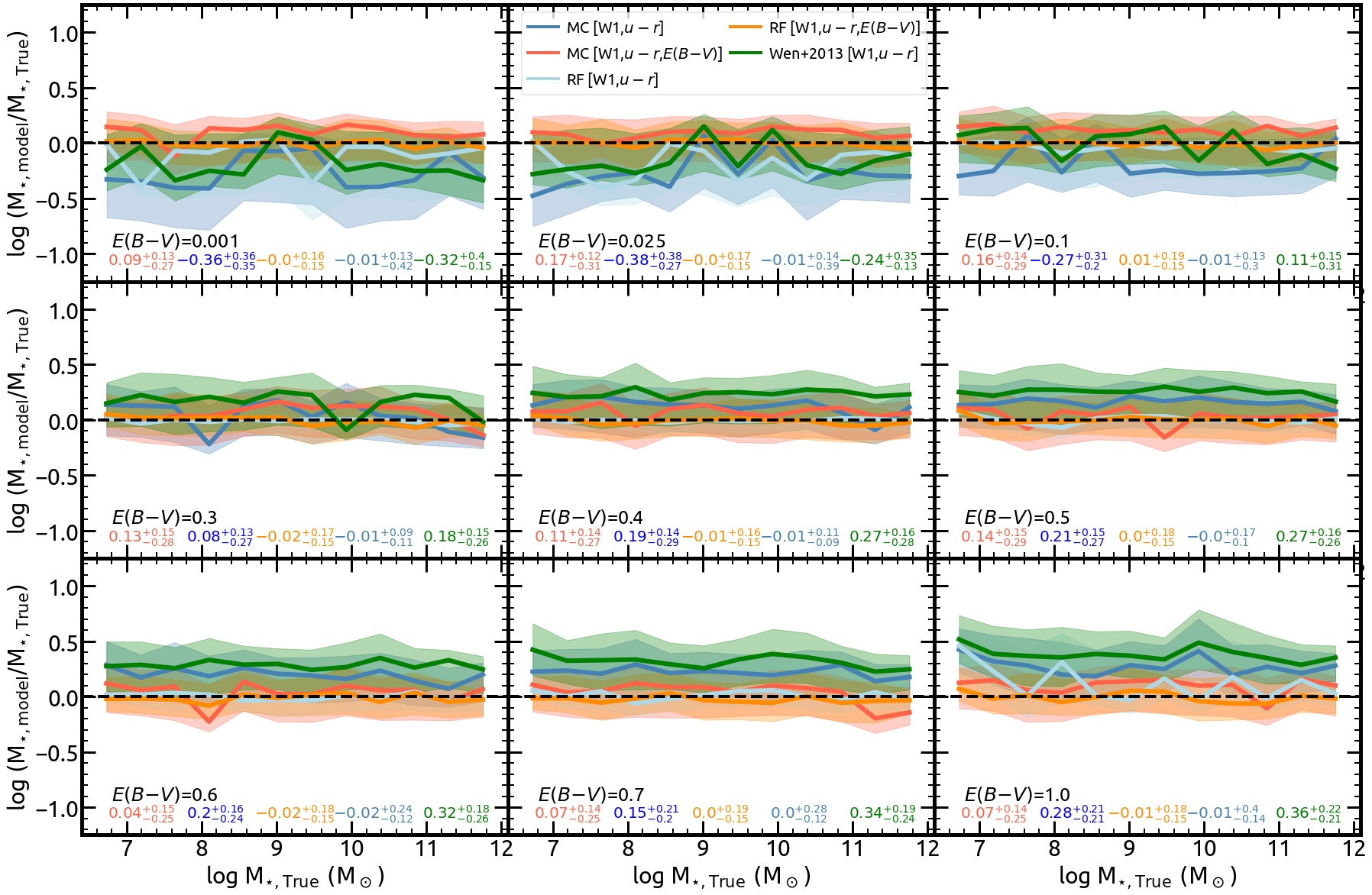}
    \caption{Comparison between the true $M_\star$ as given by the \texttt{CIGALE} output, and the best-fit models of Eq. \ref{eq:ML_ratio} (MC; Table \ref{tab:Mstar_results}), and the random forest machine learning method (RF) using the WISE band-1 and the $u{-}r$ color.
    The $y$-axis represents the logarithm of the ratio between the $M_\star$ of each model and the true $M_\star$, thus the black dashed line at zero represents equality.
    Blue and red colors show the extinction-independent, and extinction-dependent relation of Eq. \ref{eq:ML_ratio} respectively.
    With light-blue and orange colors are the extinction-independent, and extinction-dependent models based on the RF method respectively.
    The $M_\star$ calculated based on the mass-to-light conversion from \cite{2013MNRAS.433.2946W} is also shown with green color.
    The lines represent the distribution modes and the shaded areas the 68\% percentiles as a function of the log~$M_{\rm  \star~true}$.
    Each panel shows the comparison of galaxy models with one of the discrete $E(B-V)$ values, shown in the bottom left corner of each panel.
    This comparison is performed with the uniform-$E(B-V)$ sample, therefore, there is an equal number of sources for each extinction value.
    At the bottom of each panel are the modes and 68\% percentiles of the $\rm log~M_{\star model}/M_{\star, True}$ distribution for all galaxy models with the specific $E(B-V)$, shown with corresponding colors to the relations used to derive the $M_\star$.
    These comparisons are also summarized in Table \ref{tab:Mstar_dist_results}.
    }
    \label{fig:ML_ratio}
\end{figure*}

\begin{table*}[ht!]
    \renewcommand{\arraystretch}{1.75} 
    \centering
    \caption{
    The modes and 68\% percentiles for the logarithm of the ratio between the estimated $M_\star$ of the models, based on Equation \ref{eq:ML_ratio}
    using the WISE band-1 or NIR-F200W photometries and the $u{-}r$ or $g{-}r$ optical colors, over the true $M_\star$ as given by \texttt{CIGALE}. 
    Results using the \cite{2013MNRAS.433.2946W} calibrations are also given. 
    The two columns are for comparisons using the uniform-$E(B-V)$ samples, and the SDSS-matched $E(B-V)$ samples in the color range $u-r \leq 1$ (left), and $u-r >1$ (right), and $g-r \leq 0.75$ (left), and $g-r >0.75$ (right).
    }
    \begin{tabular}{l|cccc}
    \hline
    \hline
    & \multicolumn{4}{c}{$\rm <log \frac{M_{\star,model}}{M_{\star,true}}>$}\\
    \hline
      \diagbox{Model}{$E(B-V)$ distribution} & uniform & SDSS & uniform & SDSS \\
    \hline
     & \multicolumn{2}{c}{$ u-r \leq 1$} & \multicolumn{2}{c}{$u-r  > 1$}  \\
    \hline
     $[$W1, $u{-}r$]  & $ -0.02 ^{+ 0.4 }_{- 0.37 }$ &  $ -0.04 ^{+ 0.28 }_{- 0.23 }$ &  $ 0.14 ^{+ 0.16 }_{- 0.24 }$ &  $ 0.1 ^{+ 0.13 }_{- 0.25 }$\\
     $[$W1, $u{-}r$, $E(B-V)$] & $ -0.05 ^{+ 0.25 }_{- 0.19 }$ & $ -0.03 ^{+ 0.25 }_{- 0.18 }$ & $ 0.09 ^{+ 0.09 }_{- 0.29 }$ & $ 0.08 ^{+ 0.09 }_{- 0.29 }$\\
     $[$W1, $u{-}r$, IRXu] & $ -0.09 ^{+ 0.29 }_{- 0.18 }$ & $ -0.08 ^{+ 0.25 }_{- 0.19 }$ & $ 0.04 ^{+ 0.14 }_{- 0.21 }$ & $ 0.06 ^{+ 0.11 }_{- 0.22 }$\\
    \cite{2013MNRAS.433.2946W} $[$W1, $u{-}r$] & $ 0.05 ^{+ 0.43 }_{- 0.28 }$ & $ 0.17 ^{+ 0.32 }_{- 0.24 }$ & $ 0.26 ^{+ 0.14 }_{- 0.26 }$ & $ 0.26 ^{+ 0.1 }_{- 0.29 }$\\ 
     $[$F200W, $u{-}r$] & $ -0.08 ^{+ 0.34 }_{- 0.23 }$ & $ 0.15 ^{+ 0.18 }_{- 0.31 }$ & $ 0.04 ^{+ 0.15 }_{- 0.2 }$ & $ 0.05 ^{+ 0.14 }_{- 0.21 }$\\
     $[$F200W, $u{-}r$, $E(B-V)$] & $ -0.06 ^{+ 0.28 }_{- 0.2 }$ & $ 0.03 ^{+ 0.22 }_{- 0.23 }$ & $ 0.05 ^{+ 0.14 }_{- 0.21 }$ & $ 0.06 ^{+ 0.14 }_{- 0.21 }$\\
     \hline
     \hline
     & \multicolumn{2}{c}{$ g-r \leq 0.75$} & \multicolumn{2}{c}{$ g-r > 0.75$}  \\
     \hline
     $[$W1, $g{-}r$] &  $ 0.1 ^{+ 0.5 }_{- 0.22 }$ &  $ 0.18 ^{+ 0.46 }_{- 0.22 }$ &  $ 0.0 ^{+ 0.27 }_{- 0.18 }$ & $ -0.04 ^{+ 0.23 }_{- 0.18 }$\\
     $[$W1, $g{-}r$, $E(B-V)$] & $ 0.11 ^{+ 0.45 }_{- 0.23 }$ & $ 0.17 ^{+ 0.45 }_{- 0.22 }$ & $ -0.03 ^{+ 0.27 }_{- 0.16 }$ & $ -0.03 ^{+ 0.23}_{- 0.18 }$\\
     $[$W1, $g{-}r$, IRXg] & $ 0.22 ^{+ 0.32 }_{- 0.23 }$ & $ 0.21 ^{+ 0.33 }_{- 0.22 }$ & $ 0.0 ^{+ 0.19 }_{- 0.18 }$ & $ 0.05 ^{+ 0.18 }_{- 0.2 }$\\
    \cite{2013MNRAS.433.2946W} $[$W1, $g{-}r$] & $ 0.14 ^{+ 0.49 }_{- 0.22 }$ & $ 0.26 ^{+ 0.44 }_{- 0.22 }$ & $ 0.2 ^{+ 0.24 }_{- 0.21 }$ & $ 0.22 ^{+ 0.21 }_{- 0.24 }$\\      
     $[$F200W, $g{-}r$] & $ 0.21 ^{+ 0.33 }_{- 0.25 }$ & $ 0.19 ^{+ 0.33 }_{- 0.25 }$ & $ 0.05 ^{+ 0.24 }_{- 0.18 }$ & $ -0.08 ^{+ 0.25 }_{- 0.11 }$\\
     $[$F200W, $g{-}r$, $E(B-V)$] & $ 0.21 ^{+ 0.32 }_{- 0.25 }$ & $ 0.18 ^{+ 0.34 }_{- 0.26 }$ & $ -0.01 ^{+ 0.22 }_{- 0.14 }$ & $ 0.07 ^{+ 0.23 }_{- 0.18 }$\\
    \hline
    \hline
    \end{tabular}
    \label{tab:Mstar_dist_results}
\end{table*}

\section{Comparison with observations}
\label{sec:Observations_comp}

The comparisons of the SFR and mass-to-light calibrations of Equations \ref{eq:W3_W1_to_SFR} and \ref{eq:ML_ratio} show excellent results in retrieving the SFR and $M_\star$ of the mock galaxies used in our analysis.
However, in order to test these calibrations in a more realistic setting we compare them with observations of galaxies and their SFRs and stellar masses inferred from different methods.

\subsection{Comparison of the star-formation rate calibrations}
\label{sec:SFR_comp}

Figure \ref{fig:SFR_Obs_Comps} shows comparisons between SFRs based on our method of extinction-dependent and extinction-independent calibrations using Eq. \ref{eq:W3_W1_to_SFR} and the best-fit relations in Table \ref{tab:SFR_results} with
various methods including: a) SED fitting from \cite{2015ApJS..219....8C}, \cite{2018ApJ...859...11S}, or \textit{MPA-JHU} \citep{10.1111/j.1365-2966.2003.07154.x,10.1111/j.1365-2966.2004.07881.x,Tremonti_2004}, and b) those based on monochromatic WISE band-3  emission \citep{2013AJ....145....6J,2015ApJS..219....8C,2017ApJ...850...68C}.
These comparisons involve star-forming and unclassified but exclude LINER and AGN galaxies as they were classified by \textit{MPA-JHU} \citep[\texttt{BPTCLASS} $\neq$ 4 and \texttt{BPTCLASS} $\neq$ 5;][]{10.1111/j.1365-2966.2004.07881.x}.
The adopted IR fluxes for the WISE bands are from the AllWISE source catalog\footnote{http://wise2.ipac.caltech.edu/docs/release/allwise/} as given by \cite{2015ApJS..219....8C} before the photometric corrections they applied.
Distances used to calculate the luminosities were based on the redshifts provided by the \textit{MPA-JHU} catalog.
Their $E(B-V)$ was estimated through the \textit{Balmer extinction} based on the H$\alpha$ and H$\beta$ fluxes from \textit{MPA-JHU} and the conversion of \cite{2013ApJ...763..145D}.
Galaxies with $\rm SNR < 5$ in the WISE bands 1 and 3 fluxes and the $E(B-V)$ were omitted ensuring good quality data.

\begin{figure*}
    \centering
    \includegraphics[width=\textwidth]{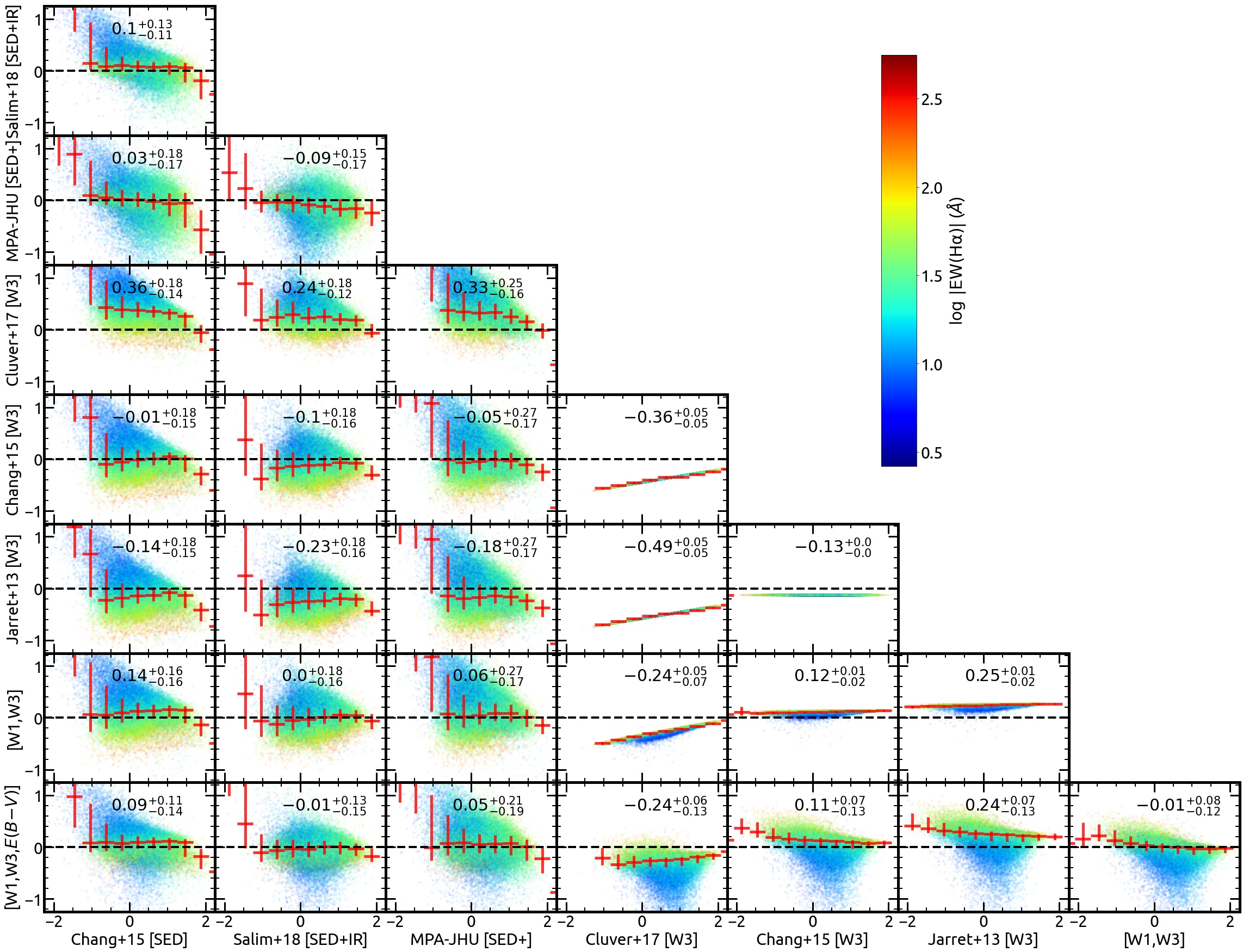}
    \caption{Comparisons between SFRs estimated through a) the extinction-dependent and extinction-independent relations of Eq. \ref{eq:W3_W1_to_SFR} (Table \ref{tab:SFR_results}) which utilize WISE bands 1 and 3, b) calibrations based only on WISE band-3 \citep{2013AJ....145....6J,2015ApJS..219....8C,2017ApJ...850...68C}, and c) based on SED fitting from \cite{2015ApJS..219....8C}, \cite{2018ApJ...859...11S} and \textit{MPA-JHU} \citep{10.1111/j.1365-2966.2003.07154.x,10.1111/j.1365-2966.2004.07881.x,Tremonti_2004}.
    In each subplot, the ordinate indicates the logarithm of the ratio ($\rm log~(SFR_y/SFR_x)$) between the compared SFR estimations, and the abscissa the logarithm of the SFR as indicated in the axes labels.
    Red error bars represent the modes and 68\% percentiles of the distribution of galaxies within bins of 
    log~$\rm SFR/M_\odot~yr^{-1} \simeq 0.4$ (12 bins within the range $\rm -2.5 < log~ SFR/M_\odot~yr^{-1} < 2.5$).
    The $\rm log~SFR_y/SFR_x$ modes and 68\% percentiles for all the galaxies compared in each panel are shown at the top of each subplot.
    The points are color-coded based on the logarithm of the absolute value of the equivalent width (EW) of the H$\alpha$ line of the galaxies.
    The black dashed line represents equality.
    }
    \label{fig:SFR_Obs_Comps}
\end{figure*}

The format of Figure \ref{fig:SFR_Obs_Comps} is similar to that of Figure \ref{fig:SFR_comp} where the ordinate indicates the logarithm of the ratio of the compared SFRs ($\rm log~SFR_y/SFR_x$), and the abscissa the logarithm of the SFRs as indicated in the axes labels.
Therefore, the equality lies at value zero of the y-axis.
The points in Figure \ref{fig:SFR_Obs_Comps} are color-coded based on the absolute value of the equivalent width of the H$\alpha$ line ($\rm EW_{H\alpha}$) of the galaxies.
The H$\alpha$ EW is strongly correlated with the sSFR of the galaxies \citep[e.g.][]{2018MNRAS.477.3014B} and it is used here as an independent method for tracing their current star-forming activity with respect to their stellar component.

The SED fitting of \cite{2015ApJS..219....8C} included optical (SDSS), and IR (WISE), but not UV photometry.
The SED fitting of \cite{2018ApJ...859...11S} was based on a SED + $L_{\rm IR}$ fitting method utilizing except the SDSS optical and GALEX UV in the SED fitting, corrections based on the 22 $\mu$m or 12 $\mu$m photometry from unWISE.
The SFR estimations from \textit{MPA-JHU} were based on the galaxies' emission lines and the method of \cite{10.1111/j.1365-2966.2003.07154.x} for galaxies within the SDSS spectroscopic fiber.
For galaxies that did not fit within the fiber, SFRs were estimated using the \textit{ugriz} photometry based on SED fitting following \cite{2007ApJS..173..267S}.

The proposed SFR calibrations of Eq. \ref{eq:W3_W1_to_SFR} show the best on-average agreement compared to the SFRs from SED fitting.
The extinction-dependent relation shows an overall slightly better agreement with the other indicators compared to the extinction-independent calibration.
Moreover, the Eq. \ref{eq:W3_W1_to_SFR} calibrations show good agreement throughout the SFR range compared to the SFRs from \cite{2018ApJ...859...11S} with on-average no large deviations in the low or high SFRs.
As expected, the extinction-dependent relation shows better behavior in the lower and higher SFR regimes.
The scatter between the Eq. \ref{eq:W3_W1_to_SFR} calibrations compared to results from SED fitting is reduced with respect to other IR-based SFRs.

The comparisons between the monochromatic IR--SFR calibrations \citep{2013AJ....145....6J,2015ApJS..219....8C,2017ApJ...850...68C} reveals some differences between them.
While the best agreement is between the calibrations of \cite{2013AJ....145....6J}, and \cite{2015ApJS..219....8C}, the largest difference is between \cite{2013AJ....145....6J}, and \cite{2017ApJ...850...68C} (about 0.5~dex on average).
As also shown from the comparisons with the mock photometry used in our analysis (Figure \ref{fig:SFR_comp}), comparing Eq. \ref{eq:W3_W1_to_SFR} with these calibrations show differences ranging from -0.25~dex \citep[compared to][]{2017ApJ...850...68C} to +0.25~dex \citep[compared to][]{2013AJ....145....6J} on average and non-linear dependence on the value of the SFR.

As revealed by the color-coding of Figure \ref{fig:SFR_Obs_Comps}, the galaxies where the Eq. \ref{eq:W3_W1_to_SFR} calibrations infer lower SFRs compared to other IR--SFR calibrations have very low EW ($\rm EW_{H\alpha} \lesssim 20$ \AA) indicating that these are passive galaxies which are mainly dominated by older SPs.
Moreover, only the extinction-dependent calibration of Eq. \ref{eq:W3_W1_to_SFR} does not show a monotonic $\rm EW_{H\alpha}$ gradient in the comparisons with the SED-fitting SFRs.
This indicates the ability of the proposed calibration to account for the contribution of older SPs and extinction in the dust heating while all other IR--SFR calibrations tend to overestimate the SFR of passive galaxies and underestimate the SFR of highly star-forming galaxies.

The SED fitting and the use of emission lines have been proven powerful methods for deriving the characteristics of galaxies.
However, not using simultaneously UV and IR photometry can lead to biases with respect to the estimation of extinction due to not covering the energy balance between the absorbed UV light, and that which is re-emitted in the IR \citep[e.g.][]{2013ApJ...768...90L}.
This can be more important in the extremes of the SFR, and $M_\star$ range, for instance in dust-free galaxies when the UV is not taken into account, or in dusty/high-SFR galaxies when the IR emission is not taken into account.
The fact that the calibrations of Eq. \ref{eq:W3_W1_to_SFR} show excellent agreement with the SFRs from SED fitting demonstrates the advantage of this method with respect to monochromatic IR SFR indicators and shows that their results can be considered robust over a wide SFR range.

\subsection{Comparison of the mass-to-light ratio calibrations}
\label{sec:Mstar_comp}

Figure \ref{fig:Corner_Mstar_obs} shows comparisons between stellar masses calculated using various methods including:
a) the extinction-dependent, and extinction-independent relations of Eq. \ref{eq:ML_ratio} which combine the WISE band 1 and the $u{-}r$, or $g{-}r$ color, b) SED fitting from \cite{2015ApJS..219....8C}, \cite{2018ApJ...859...11S}, and \textit{MPA-JHU} \citep{10.1111/j.1365-2966.2003.07154.x,10.1111/j.1365-2966.2004.07881.x,Tremonti_2004}, and c) the mass-to-light calibration of \cite{2013MNRAS.433.2946W} which is using the WISE band-1 and the $u{-}r$ color.
Similarly to Figure \ref{fig:SFR_Obs_Comps}, this comparison involves star-forming or unclassified galaxies but excludes LINER and AGN galaxies as they have been classified by \textit{MPA-JHU} (\texttt{BPTCLASS} $\neq$ 4 and \texttt{BPTCLASS} $\neq$ 5).
It involves galaxies with $\rm SNR > 5$ in the $L_{\rm W1}$, SDSS $u$, $g$, $r$ magnitudes, and $E(B-V)$ extinction metric.
WISE band fluxes were based on the AllWISE source catalog and were adopted from  \cite{2015ApJS..219....8C}.
The $E(B-V)$ was estimated through the \textit{Balmer extinction} based on the H$\alpha$ and H$\beta$ fluxes from \textit{MPA-JHU} and the conversion of \cite{2013ApJ...763..145D}.
Distances used to calculate the luminosities were based on the redshifts provided by the \textit{MPA-JHU} catalog.

\begin{figure*}
    \centering
    \includegraphics[width=\textwidth]{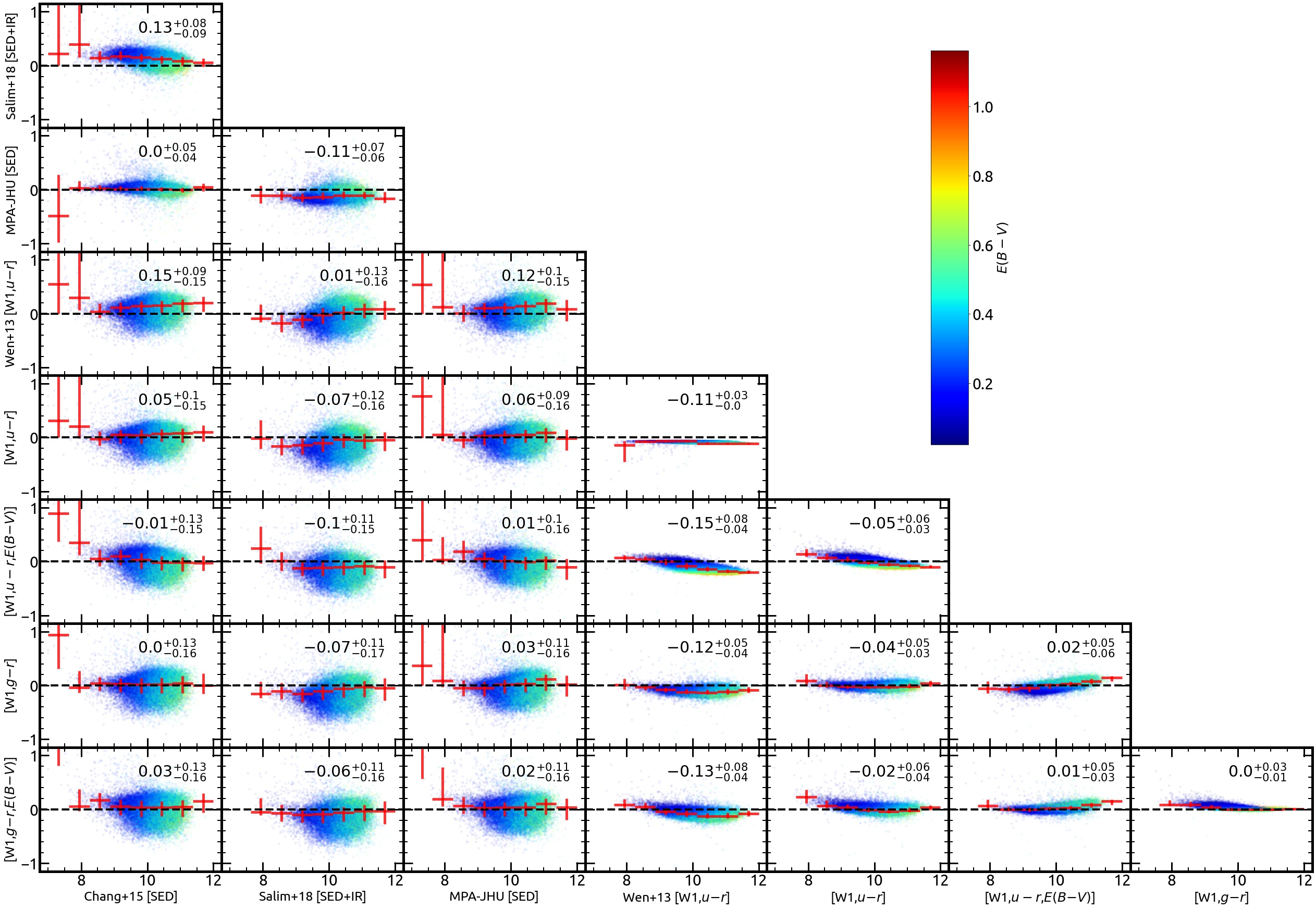}
    \caption{Comparisons between stellar masses estimated through a) the extinction-dependent, and extinction-independent relations of Eq. \ref{eq:ML_ratio} based on the MCMC fitting (Table \ref{tab:Mstar_results}) which utilize the WISE band-1 photometry and the $u{-}r$ or the $g{-}r$ color, b) the calibration of \cite{2013MNRAS.433.2946W} which uses the WISE band 1 and the $u{-}r$ color
    and c) based on SED fitting from \cite{2015ApJS..219....8C}, \cite{2018ApJ...859...11S}, and \textit{MPA-JHU} \citep{10.1111/j.1365-2966.2003.07154.x,10.1111/j.1365-2966.2004.07881.x,Tremonti_2004}.
    In each subplot, the ordinate indicates the logarithm of the ratio ($\rm log~(M_{\star,y}/M_{\star,x})$) between the compared $M_\star$ estimations, and the abscissa the logarithm of the $M_\star$ as indicated in the axes labels.
    Red error bars represent the modes and 68\% percentiles of the distribution of galaxies within bins of log~$ M_\star/M_\odot \simeq 0.6$ (8 bins within the range $7 < {\rm log}~ M_\star/M_\odot < 12$).
    The ${\rm log}~M_{\star_{\rm y}}/M_{\star_{\rm x}}$ modes and 68\% percentiles for all the galaxies compared in each panel are shown at the top right of each subplot.
    The points are color-coded based on the galaxies' nebular $E(B-V)$.
    The black dashed line represents equality.
    }
    \label{fig:Corner_Mstar_obs}
\end{figure*}

Overall, these comparisons show that calibrations using monochromatic NIR photometry and an optical color agree on average with the stellar masses based on SED fitting with a small standard deviation of about 0.1 dex.
The extinction-independent relation is in excellent agreement with the stellar masses derived using the \cite{2013MNRAS.433.2946W} calibration showing $\sim -0.1$ dex lower $M_\star$ on average without strong deviations in low or high stellar masses.
However, it shows a slightly better agreement in low and high $M_\star$ galaxies compared to results based on SED fitting, while the $u{-}r$ relation of \cite{2013MNRAS.433.2946W} shows a small overestimation for galaxies with large $M_\star$ and high extinction.

The extinction-dependent relation of Eq. \ref{eq:ML_ratio} is in good agreement with stellar masses estimated through SED fitting.
These comparisons show slightly better agreement throughout the $M_\star$ range, even for very high and low-SFR galaxies where the effect of reddening and SP age becomes gradually more important.
Moreover, as indicated by the color code of Figure \ref{fig:Corner_Mstar_obs} the extinction-dependent relation of Eq. \ref{eq:ML_ratio} tends to correct the overestimation of high extinction galaxies, and vice-versa the underestimation of low extinction galaxies, compared to extinction independent relations.
The comparison with the model SEDs, where we know a-priori the true $M_\star$ of the galaxies (Section \ref{sec:results_stellar_mass}), indicates that the inferred stellar masses based on the extinction-dependent relation of Eq. \ref{eq:ML_ratio} and on SED fitting are closer to the correct $M_\star$.
These comparisons also demonstrate the importance of taking extinction into account in the estimation of the stellar mass.

\section{Conclusions}
\label{sec:conclusions}

We used the SED fitting code \texttt{CIGALE} to create a large sample of mock galaxy SEDs that covered a wide range of SFRs, stellar masses, and ISM conditions. 
We compared the SFR and $M_\star$ of the model galaxies with their IR luminosity for photometric bands of WISE and the JWST MIRI and NIRCam instruments. 
This analysis showed that there is a strong dependence of the IR tracers of galaxies' SFR and $M_\star$ on the age of their SPs, and their extinction.
However, the combined use of IR bands that trace SPs of different ages can provide more robust calibrations and minimize the biases and the scatter they introduce in the SFR, and $M_\star$ calibrations. 
This is particularly important for low-SFR galaxies where the contribution of old SPs can be significant when measuring their SFR, and for high-SFR galaxies where the contribution of dust emission can bias the $M_\star$ measurements.
The addition of extinction-dependent calibrations also offers more reliable SFR, and $M_\star$ estimations with less scatter which are particularly better for low-extinction/dust-free (e.g. dwarf) galaxies.

In summary, this work provides extinction-dependent and extinction-independent calibrations and quantification of their scatter for measuring the:
   \begin{enumerate}
      \item SFR based on the WISE band-3 and band-1 utilizing as extinction indicators the $E(B{-}V)$, the $L_{\rm W4}/L_{u}$, and the $L_{\rm W4}/L_{g}$ ratio (Eq. \ref{eq:W3_W1_to_SFR}, Table \ref{tab:SFR_results})
      \item SFR based on the JWST NIR-F200W and MIRI-F2100W bands utilizing as extinction indicator the $E(B{-}V)$ (Eq. \ref{eq:W3_W1_to_SFR}, Table \ref{tab:SFR_results})
      \item mass-to-light ratio based on the WISE band-1 and the $u{-}r$ or the $g{-}r$ color utilizing as extinction indicators the $E(B{-}V)$, the $L_{\rm W4}/L_{u}$, and the $L_{\rm W4}/L_{g}$ ratio (Eq. \ref{eq:ML_ratio}, Table \ref{tab:Mstar_results})
      \item mass-to-light ratio based on the JWST NIR-F200W and the $u{-}r$ or $g{-}r$ color utilizing as extinction indicator the $E(B{-}V)$ (Eq. \ref{eq:ML_ratio}, Table \ref{tab:Mstar_results}).
   \end{enumerate}
   
The comparisons with both modeled and observed samples of galaxies show that the proposed calibrations offer robust SFR, and $M_\star$ estimations for a wide range of these values, minimizing the scatter. 
The random forest analysis
yields similar results to the MCMC method showing that the provided calibrations are method independent.

\begin{acknowledgements}

The authors thank the anonymous referee for providing comments and suggestions that improved the clarity of this work.
KK is supported by the project "Support of the international collaboration in astronomy (Asu mobility)" with the number: CZ.02.2.69/0.0/0.0/18\_053/0016972. Asu mobility is co-financed by the European Union.
KK and AZ acknowledge funding from the European Research Council under the European Union's Seventh Framework Programme (FP/2007-2013)/ERC Grant Agreement n. 617001 (A-BINGOS). 
This project has received funding from the European Union's Horizon 2020 research and innovation programme under the Marie Sklodowska-Curie RISE action, grant agreements No 691164 (ASTROSTAT), and No 873089 (ASTROSTAT-II).
EK acknowledges support from the Public Investments Program through a Matching Funds grant to the IA-FORTH.
SS is supported through NASA awards NNX12AE06G and 80NSSC20K0440.
JS acknowledges financial support from the Czech Science Foundation under Project No. 22-22643S.
This research has made use of: 
(a) data products from the Wide-field Infrared Survey Explorer (WISE), which is a joint project of the University of California, Los Angeles, and JPL, California Institute of Technology, funded by NASA; 
(b) observations made with the Spitzer Space Telescope, which was operated by JPL, California Institute of Technology under a contract with NASA; 
(c) the NASA/IPAC Extragalactic Database (NED), which is operated by the Jet Propulsion Laboratory (JPL), California Institute of Technology, under contract with NASA; 
(d) the NASA/IPAC Infrared Science Archive (IRSA), which is funded by NASA and operated by the California Institute of Technology; 

\end{acknowledgements}

\bibliographystyle{aa} 
\bibliography{citations.bib} 


\section*{Appendix A}
\label{sec:Appendix}

\numberwithin{table}{section}
\setcounter{table}{0}
\renewcommand{\thetable}{A\arabic{table}}

Table \ref{tab:pcigale_ini} shows the configuration of the \texttt{CIGALE} SED creation modules. See text of \cite{2019A&A...622A.103B} for a detailed description of the parameters.

\begin{table*}[ht!]
    \centering
    \caption{Configuration of the \texttt{CIGALE} SED creation modules.}
    \begin{tabular}{l|c|r}
       \toprule
       Module & sfhdelayed   \\
       \hline
       tau\_main & 500, 1000, 2000, 4000, 6000, 8000 & e-folding time of the main SP model in Myr\\
       age\_main & 8000, 13000 & Age of the main SP in the galaxy in Myr \\
       tau\_burst & 10, 25, 50, 100, 250, 500, 1000, 2000 & e-folding time of the late starburst population\\
       age\_burst & 3000, 2000, 1000, 500, 250, 100, 50, 25, 10 & Age of the late burst in Myr\\
       f\_burst & 0.25, 0.1, 0.05, 0.01, 0.001, 0.0001, 0.0 & Mass fraction of the late burst population\\
       sfr\_A & 10,000 & Multiplicative factor controlling the SFR\\
       normalise & False & Normalise SFH to produce one solar mass\\
       \hline
       \hline
       Module & bc03\\
       \hline
       imf & 0 & Salpeter initial mass function\\
       metallicity & 0.0001, 0.0004, 0.004, 0.008, 0.02, 0.05 & Z\\
       separation\_age & 10 & Young and old SPs separation age in Myr\\ 
       \hline
       \hline
       Module & nebular \\
       \hline
       logU & -1.0, -4.0 & Ionisation parameter\\
       f\_esc & 0.1 & Fraction of escaping $Ly$ continuum photons\\
       f\_dust & 0.1 & Fraction of absorbed $Ly$ continuum photons\\
       lines\_width & 300.0 & Line width (km/s)\\
       emission & True & Include nebular emission\\
       \hline
       \hline
       Module & dustatt\_modified\_starburst \\
       \hline
       E\_BV\_lines & 0.001, 0.05, 0.1, 0.15, 0.2, 0.25, 0.3 & color excess of the nebular lines\\
       & 0.35, 0.4, 0.45, 0.5, 0.55, 0.6, 0.65\\
       & 0.7, 0.75, 0.8, 0.85, 0.9, 0.95, 1.0 & \\ 
       E\_BV\_factor &  0.3 & Reduction factor applied on E\_BV\_lines.\\
       uv\_bump\_wavelength & $217.5$ & Central wavelength of the UV bump (nm)\\
       uv\_bump\_width & $35.0$ & Width (FWHM) of the UV bump (nm)\\
       uv\_bump\_amplitude & $0.0$ & Amplitude of the UV bump \\
       powerlaw\_slope & $-1.5$, $-0.5$, $0.0$, $0.5$ & Modifying slope $\delta$\\
       Ext\_law\_emission\_lines & 1 & 1 corresponds to MW extinction law\\
       $ R_{\rm V}$ & 3.1 & $A_{\rm V}/E(B-V)$\\
       \hline
       \hline
       Module & dale2014 \\
       \hline
       fracAGN & 0.0 & AGN fraction \\
       alpha & 2.0 & Alpha slope \\
       \hline
       \hline
       Module & restframe\_parameters\\
       \hline
       beta\_calz94 & True & UV slope as in Calzetti et al. (1994)\\
       D4000 & True & As in Balogh et al. (1999)\\
       IRX & True & based on GALEX FUV and dust luminosity\\
       \hline
       \hline
       Module & redshifting \\
       \hline
       redshift & 0 \\
    \bottomrule
    \end{tabular}
    \label{tab:pcigale_ini}
\end{table*}



   

\end{document}